\documentclass[12pt]{amsart}
\usepackage{feynmp}
\usepackage{latexsym}
\usepackage{amsfonts}
\usepackage{amssymb}
\usepackage{amscd}
\usepackage{amsbsy}
\usepackage{amsmath}
\usepackage[dvips]{graphicx}
\usepackage{epsfig}    
\usepackage[all]{xy}

\def\cf{{\it cf.\ }}
\def\rhs{{\it r.h.s.\ }}

\def\End{\mathop{{\rm End}}\nolimits}

\def\Hom{\mathop{{\rm Hom}}\nolimits}
\def\Ext{\mathop{{\rm Ext}}\nolimits}

\def\im{\mathop{{\rm im}}\nolimits}
\def\Tr{\mathop{{\rm Tr}}\nolimits}
\def\STr{\mathop{{\rm STr}}\nolimits}

\def\deg{ \mathop{{\rm deg}}\nolimits }
\def\p{^{\prime}}

\def\prosign{\mathop{{\rm sign}}\nolimits}
\def\sign#1{\prosign\left(#1\right)}

\def\rank{\mathop{{\rm rank}}\nolimits}

\def\del{ \partial }

\def\leg#1{ \{ #1 \}_K }

\def\End{\mathop{{\rm End}}\nolimits}

\def\Pexp{\mathop{{\rm Pexp}}\nolimits}

\def\lrbc#1{ \left( #1 \right) }
\def\lrbs#1{ \left[ #1 \right] }

\def\lrBigc#1{ \Big( #1 \Big) }

\def\corr#1{ \langle #1 \rangle }
\def\corri#1#2{ \corr{#1}_{#2} }

\def\xG{ \mfG }

\def\xmap#1#2#3{ #1\colon #2\longrightarrow #3 }

\parskip=\medskipamount                 
\thicklines                     
\newcommand{\bimn}[7]{\bibitem{#1}#2,
{\em #3}, { #4}\hspace{0.25em}{\bf
#5}\hspace{0.25em}(#6)\hspace{0.25em}{#7}.}

\def\inbar{\vrule height1.5ex width.4pt depth0pt}
\def\IC{\relax\,\hbox{$\inbar\kern-.3em{\rm C}$}}
\def\IN{\relax{\rm I\kern-.18em N}}
\def\IQ{\relax\,\hbox{$\inbar\kern-.3em{\rm Q}$}}
\def\IR{\relax{\rm I\kern-.18em R}}
\def\ZZ{\relax{\sf Z\kern-.4em Z}}

   \def\cD{{\cal D}}

  \def\cL{{\cal L}}

\catcode`\@=11

\newif\if@fewtab\@fewtabtrue

\catcode`\@=11

\newif\if@fewtab\@fewtabtrue

{\count255=\time\divide\count255 by 60
\xdef\hourmin{\number\count255} \multiply\count255
by-60\advance\count255 by\time
\xdef\hourmin{\hourmin:\ifnum\count255<10 0\fi\the\count255}}
\def\ps@draft{\let\@mkboth\@gobbletwo
    \def\@oddhead{}
    \def\@oddfoot
      {\hbox to 7 cm{\footnotesize {\em Draft of \jobname:} \draftdate
       \hfil}\hskip -7cm\hfil\rm\thepage \hfil}
    \def\@evenhead{}\let\@evenfoot\@oddfoot}

\def\ceqno{\global\@fewtabfalse
    \ifcase\@eqcnt \def\@tempa{& & &}\or \def\@tempa{& &}
      \or \def\@tempa{&}
      \or\def\@tempa{}\fi\@tempa
{\rm(\theequation)}}

\def\aeqno#1{\global\@fewtabfalse
    \ifcase\@eqcnt \def\@tempa{& & &}\or \def\@tempa{& &}
      \or \def\@tempa{&}
      \or\def\@tempa{}\fi\@tempa
{\rm(\theequation,#1)}}

\def\label#1{\ifnum\draftcontrol=1
 \global\def\draftnote{$\scriptstyle #1$}\fi
 \@bsphack\if@filesw {\let\thepage\relax
   \def\protect{\noexpand\noexpand\noexpand}%
\xdef\@gtempa{\write\@auxout{\string
      \newlabel{#1}{{\@currentlabel}{\thepage}}}}}\@gtempa
   \if@nobreak \ifvmode\nobreak\fi\fi\fi
  \@esphack}

\def\alabel#1#2{\label{#1}\global\@fewtabfalse
    \ifcase\@eqcnt \def\@tempa{& & &}\or \def\@tempa{& &}
      \or \def\@tempa{&}
      \or\def\@tempa{}\fi\@tempa
{\hbox to 3cm{\phantom{\rm(\theequation,#2)} \draftnote
\hfil}\hskip -3cm {\rm(\theequation,#2)}}}

\def\clabel#1{\label{#1}\global\@fewtabfalse
    \ifcase\@eqcnt \def\@tempa{& & &}\or \def\@tempa{& &}
      \or \def\@tempa{&}
      \or\def\@tempa{}\fi\@tempa
{\hbox to 3cm{\phantom{\rm(\theequation)} \draftnote \hfil}\hskip
-3cm{\rm(\theequation)}}}

\def\eqnarray{\def\draftnote{{}}\global\@fewtabtrue
\stepcounter{equation}\let\@currentlabel=\theequation
\global\@eqnswtrue
\global\@eqcnt\z@\tabskip\@centering\let\\=\@eqncr
$$\halign to \displaywidth\bgroup\@eqnsel\hskip\@centering\@eqcnt\z@
  $\displaystyle\tabskip\z@{##}$&\global\@eqcnt\@ne
  \hskip 1\arraycolsep \hfil$\displaystyle{##}$\hfil
  &\global\@eqcnt\tw@ \hskip 1\arraycolsep
$\displaystyle\tabskip\z@{##}$ \hfil
\tabskip\@centering&\global\@eqcnt\thr@@\llap{##}\tabskip\z@ \cr}

\def\endeqnarray{\@@eqncr\egroup
      \global\advance\c@equation\m@ne$$\global\@ignoretrue}

\def\@eqnnum{\hbox to 3cm{\phantom{\rm(\theequation)} \draftnote
                         \hfil}\hskip -3cm {\rm(\theequation)}}

\def\@@eqncr{\let\@tempa\relax
    \ifcase\@eqcnt \def\@tempa{& & &}\or \def\@tempa{& &}
      \or \def\@tempa{&}
      \or\def\@tempa{}
\fi\@tempa \if@eqnsw \if@fewtab\@eqnnum\fi
\stepcounter{equation}\fi\global
\@eqnswtrue\global\@eqcnt\z@\global\@fewtabtrue\cr}

\def\draftcite#1{\ifnum\draftcontrol=1#1\else{}\fi}

\def\@lbibitem[#1]#2{\item{}\hskip -3cm \hbox to 2cm
{\hfil$\scriptstyle\draftcite{#2}$}\hskip
1cm[\@biblabel{#1}]\if@filesw
     {\def\protect##1{\string ##1\space}\immediate
      \write\@auxout{\string\bibcite{#2}{#1}}}\fi\ignorespaces}

\def\@bibitem#1{\item\hskip -3cm \hbox to 2cm
{\hfil $\scriptstyle\draftcite{#1}$}\hskip 1cm \if@filesw
\immediate\write\@auxout
       {\string\bibcite{#1}{\the\value{\@listctr}}}\fi\ignorespaces}

\def\draftdate{\number\month/\number\day/\number\year\ \ \ \hourmin }
 \global\def\draftcontrol{0}
\catcode`\@=12

\def\theequation{{\thesection.\arabic{equation}}}

\def\qq{\begin{eqnarray}}
\def\qqq{\end{eqnarray}}
\def\rx#1{~(\ref{#1})}

\def\ex#1{eq.\hspace*{-3pt}\rx{#1}}
\def\eex#1{eqs.\hspace*{-3pt}\rx{#1}}
\def\cx#1{~\cite{#1}}
\def\rw#1{~\ref{#1}}

\newlength{\shiftwidth}
\def\shift#1{&&\hbox to \shiftwidth{\hfill $\displaystyle#1$}}
\newlength{\sshiftwidth}
\def\sshift#1{\lefteqn{\hbox to
\sshiftwidth{\hfill$\displaystyle#1$}}}

\def\qbezier{\bezier{120}}
\def\DottedCircle{
\bezier{4}(0.966,-0.259)(1.04,0)(0.966,0.259)
\bezier{4}(0.966,0.259)(0.897,0.518)(0.707,0.707)
\bezier{4}(0.707,0.707)(0.518,0.897)(0.259,0.966)
\bezier{4}(0.259,0.966)(0,1.04)(-0.259,0.966)
\bezier{4}(-0.259,0.966)(-0.518,0.897)(-0.707,0.707)
\bezier{4}(-0.707,0.707)(-0.897,0.518)(-0.966,0.259)
\bezier{4}(-0.966,0.259)(-1.04,0)(-0.966,-0.259)
\bezier{4}(-0.966,-0.259)(-0.897,-0.518)(-0.707,-0.707)
\bezier{4}(-0.707,-0.707)(-0.518,-0.897)(-0.259,-0.966)
\bezier{4}(-0.259,-0.966)(0,-1.04)(0.259,-0.966)
\bezier{4}(0.259,-0.966)(0.518,-0.897)(0.707,-0.707)
\bezier{4}(0.707,-0.707)(0.897,-0.518)(0.966,-0.259) }
\def\Endpoint[#1]{
\ifcase#1 \put(1,0){\circle*{0.15}}
\or\put(0.866,0.5){\circle*{0.15}}
\or\put(0.5,0.866){\circle*{0.15}} \or\put(0,1){\circle*{0.15}}
\or\put(-0.5,0.866){\circle*{0.15}}
\or\put(-0.866,0.5){\circle*{0.15}} \or\put(-1,0){\circle*{0.15}}
\or\put(-0.866,-0.5){\circle*{0.15}}
\or\put(-0.5,-0.866){\circle*{0.15}} \or\put(0,-1){\circle*{0.15}}
\or\put(0.5,-0.866){\circle*{0.15}}
\or\put(0.866,-0.5){\circle*{0.15}} \fi}
\def\Arc[#1]{
\thicklines         
\ifcase#1 \bezier{25}(0.966,-0.259)(1.04,0)(0.966,0.259) \or
\bezier{25}(0.966,0.259)(0.897,0.518)(0.707,0.707) \or
\bezier{25}(0.707,0.707)(0.518,0.897)(0.259,0.966) \or
\bezier{25}(0.259,0.966)(0,1.04)(-0.259,0.966) \or
\bezier{25}(-0.259,0.966)(-0.518,0.897)(-0.707,0.707) \or
\bezier{25}(-0.707,0.707)(-0.897,0.518)(-0.966,0.259) \or
\bezier{25}(-0.966,0.259)(-1.04,0)(-0.966,-0.259) \or
\bezier{25}(-0.966,-0.259)(-0.897,-0.518)(-0.707,-0.707) \or
\bezier{25}(-0.707,-0.707)(-0.518,-0.897)(-0.259,-0.966) \or
\bezier{25}(-0.259,-0.966)(0,-1.04)(0.259,-0.966) \or
\bezier{25}(0.259,-0.966)(0.518,-0.897)(0.707,-0.707) \or
\bezier{25}(0.707,-0.707)(0.897,-0.518)(0.966,-0.259) \fi}
\def\DottedArc[#1]{
\ifcase#1 \bezier{4}(0.966,-0.259)(1.04,0)(0.966,0.259) \or
\bezier{4}(0.966,0.259)(0.897,0.518)(0.707,0.707) \or
\bezier{4}(0.707,0.707)(0.518,0.897)(0.259,0.966) \or
\bezier{4}(0.259,0.966)(0,1.04)(-0.259,0.966) \or
\bezier{4}(-0.259,0.966)(-0.518,0.897)(-0.707,0.707) \or
\bezier{4}(-0.707,0.707)(-0.897,0.518)(-0.966,0.259) \or
\bezier{4}(-0.966,0.259)(-1.04,0)(-0.966,-0.259) \or
\bezier{4}(-0.966,-0.259)(-0.897,-0.518)(-0.707,-0.707) \or
\bezier{4}(-0.707,-0.707)(-0.518,-0.897)(-0.259,-0.966) \or
\bezier{4}(-0.259,-0.966)(0,-1.04)(0.259,-0.966) \or
\bezier{4}(0.259,-0.966)(0.518,-0.897)(0.707,-0.707) \or
\bezier{4}(0.707,-0.707)(0.897,-0.518)(0.966,-0.259) \fi}
\def\Chord[#1,#2]{
\thinlines \ifnum#1>#2\Chord[#2,#1] \else\ifnum#1<#2 \ifcase#1
\ifcase#2 \or\qbezier(1,0)(0.516,0.138)(0.866,0.5)
\or\qbezier(1,0)(0.45,0.26)(0.5,0.866)
\or\qbezier(1,0)(0.327,0.327)(0,1)
\or\qbezier(1,0)(0.179,0.311)(-0.5,0.866)
\or\qbezier(1,0)(0.0536,0.2)(-0.866,0.5) \or\put(1, 0){\line(-2,
0){2}} \or\qbezier(1,0)(0.0536,-0.2)(-0.866,-0.5)
\or\qbezier(1,0)(0.179,-0.311)(-0.5,-0.866)
\or\qbezier(1,0)(0.327,-0.327)(0,-1)
\or\qbezier(1,0)(0.45,-0.26)(0.5,-0.866)
\or\qbezier(1,0)(0.516,-0.138)(0.866,-0.5) \fi \or\ifcase#2\or
\or\qbezier(0.866,0.5)(0.378,0.378)(0.5,0.866)
\or\qbezier(0.866,0.5)(0.26,0.45)(0,1)
\or\qbezier(0.866,0.5)(0.12,0.446)(-0.5,0.866)
\or\qbezier(0.866,0.5)(0,0.359)(-0.866,0.5)
\or\qbezier(0.866,0.5)(-0.0536,0.2)(-1,0) \or\put(0.866,
0.5){\line(-5, -3){1.73}}
\or\qbezier(0.866,0.5)(0.146,-0.146)(-0.5,-0.866)
\or\qbezier(0.866,0.5)(0.311,-0.179)(0,-1)
\or\qbezier(0.866,0.5)(0.446,-0.12)(0.5,-0.866)
\or\qbezier(0.866,0.5)(0.52,0)(0.866,-0.5) \fi \or\ifcase#2\or\or
\or\qbezier(0.5,0.866)(0.138,0.516)(0,1)
\or\qbezier(0.5,0.866)(0,0.52)(-0.5,0.866)
\or\qbezier(0.5,0.866)(-0.12,0.446)(-0.866,0.5)
\or\qbezier(0.5,0.866)(-0.179,0.311)(-1,0)
\or\qbezier(0.5,0.866)(-0.146,0.146)(-0.866,-0.5) \or\put(0.5,
0.866){\line(-3, -5){1}} \or\qbezier(0.5,0.866)(0.2,-0.0536)(0,-1)
\or\qbezier(0.5,0.866)(0.359,0)(0.5,-0.866)
\or\qbezier(0.5,0.866)(0.446,0.12)(0.866,-0.5) \fi
\or\ifcase#2\or\or\or \or\qbezier(0,1.)(-0.138,0.516)(-0.5,0.866)
\or\qbezier(0,1.)(-0.26,0.45)(-0.866,0.5)
\or\qbezier(0,1.)(-0.327,0.327)(-1,0)
\or\qbezier(0,1.)(-0.311,0.179)(-0.866,-0.5)
\or\qbezier(0,1.)(-0.2,0.0536)(-0.5,-0.866) \or\put(0, 1){\line(0,
-2){2}} \or\qbezier(0,1.)(0.2,0.0536)(0.5,-0.866)
\or\qbezier(0,1.)(0.311,0.179)(0.866,-0.5) \fi
\or\ifcase#2\or\or\or\or
\or\qbezier(-0.5,0.866)(-0.378,0.378)(-0.866,0.5)
\or\qbezier(-0.5,0.866)(-0.45,0.26)(-1,0)
\or\qbezier(-0.5,0.866)(-0.446,0.12)(-0.866,-0.5)
\or\qbezier(-0.5,0.866)(-0.359,0)(-0.5,-0.866)
\or\qbezier(-0.5,0.866)(-0.2,-0.0536)(0,-1) \or\put(-0.5,
0.866){\line(3, -5){1}}
\or\qbezier(-0.5,0.866)(0.146,0.146)(0.866,-0.5) \fi
\or\ifcase#2\or\or\or\or\or
\or\qbezier(-0.866,0.5)(-0.516,0.138)(-1,0)
\or\qbezier(-0.866,0.5)(-0.52,0)(-0.866,-0.5)
\or\qbezier(-0.866,0.5)(-0.446,-0.12)(-0.5,-0.866)
\or\qbezier(-0.866,0.5)(-0.311,-0.179)(0,-1)
\or\qbezier(-0.866,0.5)(-0.146,-0.146)(0.5,-0.866) \or\put(-0.866,
0.5){\line(5, -3){1.73}} \fi \or\ifcase#2\or\or\or\or\or\or
\or\qbezier(-1,0)(-0.516,-0.138)(-0.866,-0.5)
\or\qbezier(-1,0)(-0.45,-0.26)(-0.5,-0.866)
\or\qbezier(-1,0)(-0.327,-0.327)(0,-1)
\or\qbezier(-1,0)(-0.179,-0.311)(0.5,-0.866)
\or\qbezier(-1,0)(-0.0536,-0.2)(0.866,-0.5) \fi
\or\ifcase#2\or\or\or\or\or\or\or
\or\qbezier(-0.866,-0.5)(-0.378,-0.378)(-0.5,-0.866)
\or\qbezier(-0.866,-0.5)(-0.26,-0.45)(0,-1)
\or\qbezier(-0.866,-0.5)(-0.12,-0.446)(0.5,-0.866)
\or\qbezier(-0.866,-0.5)(0,-0.359)(0.866,-0.5) \fi
\or\ifcase#2\or\or\or\or\or\or\or\or
\or\qbezier(-0.5,-0.866)(-0.138,-0.516)(0,-1)
\or\qbezier(-0.5,-0.866)(0,-0.52)(0.5,-0.866)
\or\qbezier(-0.5,-0.866)(0.12,-0.446)(0.866,-0.5) \fi
\or\ifcase#2\or\or\or\or\or\or\or\or\or
\or\qbezier(0,-1.)(0.138,-0.516)(0.5,-0.866)
\or\qbezier(0,-1.)(0.26,-0.45)(0.866,-0.5) \fi
\or\ifcase#2\or\or\or\or\or\or\or\or\or\or
\or\qbezier(0.5,-0.866)(0.378,-0.378)(0.866,-0.5) \fi\fi\fi\fi}
\def\FullChord[#1,#2]{
\Endpoint[#1] \Endpoint[#2] \Arc[#1] \Arc[#2] \Chord[#1,#2] }
\def\EndChord[#1,#2]{
\Endpoint[#1] \Endpoint[#2] \Chord[#1,#2] }
\def\Picture#1{
\begin{picture}(2,1)(-1,-0.167)
#1
\end{picture}
}
\def\DottedChordDiagram[#1,#2]{
\Picture{\DottedCircle \FullChord[#1,#2]} }
\def\ZZ{ \mathbb{Z} }
\def\IQ{ \mathbb{Q} }
\def\IC{ \mathbb{C} }
\def\IR{ \mathbb{R} }

\def\IH{ \mathbb{H} }

\def\cD{ \mathcal{D} }

\def\cL{ \mathcal{L} }

\def\hlfv{ {1\over 2} }

\def\xJ{ J }
\def\Jv#1#2#3{ J_{#1}(#2;#3) }

\def\xP{ P }

\def\Grsm{Grassmannian}

\def\Am{A-model}

\def\TAm{Topological \Am}
\def\TAms{\TAm s}

\def\ws{world-sheet}

\def\sgr{seam graph}

\def\QFT{QFT}
\def\QFTs{\QFT s}

\def\Grv#1{ \Grl_{#1} }

\def\Grmn{ \Grvn{m} }

\def\Grmin{ \Grvn{\msi} }

\def\ICv#1{ \IC^{#1} }
\def\ICmi{ \ICv{\msi} }

\def\ICw#1{ \IC[#1] }
\def\ICbphiye{ \ICw{\bphiye} }

\def\zb{ \bar{z} }

\def\yed{ \epsilon }
\def\yeds{ \xdu{\yed} }

\def\puncture{puncture}
\def\punctures{\puncture s}

\def\adjacent{adjacent}
\def\qft{quantum field theory}

\def\ws{world-sheet}
\def\bulk{bulk}
\def\foam{foam}
\def\wsf{\ws\ \foam}
\def\wsfs{\wsf s}
\def\dwsf{\dcr\ \wsf}
\def\LG{Landau-Ginzburg}
\def\tLG{topological \LG}
\def\tLGm{\tLG\ model}
\def\LGt{\LG\ theory}

\def\tLGms{\tLGm s}
\def\aLG{LG}
\def\xTLGT{topological \aLG\ theory}
\def\xTLGTs{topological \aLG\ theories}

\def\pf{partition function}

\def\CW{\textit{CW}}
\def\tdm{2-dimensional}
\def\TQFT{T\QFT}
\def\Amdl{A-model}
\def\tAmdl{topological \Amdl}
\def\tAmdls{\tAmdl s}
\def\smdl{$\sigma$-model}
\def\Grmn{Grassmannian}
\def\Grmns{\Grmn s}
\def\seam{seam}
\def\smgr{\seam\ graph}
\def\smgrvr{\smgr\ vertex}
\def\smvr{\seam\ vertex}
\def\smvrs{\seam\ vertices}
\def\smed{\seam\ edge}
\def\smeds{\smed s}
\def\Lgr{Lagrangian}
\def\xLgr{Lagrangian}
\def\bLgr{\bulk\ \xLgr}
\def\xLgr{\Lgr}
\def\lgsm{\Lgr\ submanifold}
\def\lgsms{\lgsm s}
\def\Hlb{Hilbert}
\def\Hlbs{\Hlb\ space}
\def\Hlbss{\Hlbs s}
\def\pf{partition function}
\def\crr{correlator}
\def\crrs{\crr s}
\def\ts{target space}
\def\BRST{BRST}
\def\BRSTinv{\BRST-invariance}

\def\mf{matrix factorization}
\def\mfs{\mf s}
\def\trpl{triple}

\def\Wl{Wilson line}
\def\Wls{\Wl s}
\def\ntw{network}
\def\Wntw{Wilson \ntw}
\def\sc{connection}
\def\pigr{path integrand}
\def\Jc{Jacobi}
\def\Jca{\Jc\ algebra}
\def\Jcas{\Jca s}
\def\locop{local operator}
\def\locops{\locop s}
\def\bstate{boundary state}
\def\bstateop{\bstate\ operator}
\def\bstateops{\bstateop s}

\def\HOMFLY{HOMFLY}

\def\nghb{neighborhood}
\def\snghb{small \nghb}
\def\snghbs{\snghb s}

\def\tw{twisted}
\def\twdif{\tw\ differential}
\def\twdifs{\twdif s}

\def\scom{super-commute}
\def\scoms{\scom s}
\def\scommu{super-commutator}

\def\scommm#1#2{ \lrbs{#1,#2}_{\mathrm{s}} }

\def\ccl{cycle}

\def\Fr{Frobenius}
\def\Frt{\Fr\ trace}

\def\sp{super-potential}
\def\sps{\sp s}
\def\st{super-trace}
\def\sts{\st s}

\def\preexp{pre-exponential}

\def\tcategory{2-category}

\def\ocategory{1-category}

\def\tramp{transition amplitude}
\def\tramps{\tramp s}

\def\gluing{gluing}

\def\lcsp{local space}
\def\lcspsc{\lcsp\ section}
\def\spsc{space section}
\def\spscs{\spsc s}
\def\lcspscs{\lcspsc s}
\def\dlcspsc{\dcr\ \lcspsc}
\def\dlcspscs{\dlcspsc s}
\def\lcgr{local graph}
\def\lcgrs{\lcgr s}
\def\prtl{partial}

\def\plcgr{\prtl\ \lcgr}
\def\plcgrs{\plcgr s}
\def\dcr{decorated}
\def\dcrtn{decoration}
\def\dlcgr{\dcr\ \lcgr}

\def\dplgr{\dcr\ \plcgr}
\def\dual{dual}
\def\conjugate{conjugate}
\def\conjugated{\conjugate d}

\def\tlsc{time-like section}
\def\tlscs{\tlsc s}

\def\leg{leg}
\def\legs{\leg s}

\def\sl{slice}

\def\boundary{boundary}

\def\bvertices{\boundary\ vertices}

\def\bphi{ \boldsymbol{\phi} }
\def\bbphi{ \overline{\bphi} }
\def\bphiv#1{ \bphi_{#1} }
\def\bphii{ \bphiv{\xi} }
\def\bphim{ \bphiv{\yym} }

\def\bphimi{ \bphiv{\yymi} }
\def\bphie{ \bphiv{\xe} }
\def\bphiye{ \bphiv{\yed} }

\def\bphia{ \bphiv{\aplg} }

\def\edLa{ \yed\in\La }

\def\boeta{ \boldsymbol{\eta} }
\def\btheta{ \boldsymbol{\theta} }
\def\brho{ \boldsymbol{\rho} }
\def\bdel{ \boldsymbol{\del} }

\def\seqllv#1#2{#1_1,\ldots,{#1}_{#2} }

\def\SU{\mathrm{SU}}

\def\wsS{ \Sigma }
\def\wsSb{ \bar{\wsS} }
\def\dwsS{ \del\wsS }
\def\wsSv#1{ \wsS_{#1} }
\def\wsSbv#1{ \wsSb_{#1} }
\def\wsSi{ \wsSv{i} }
\def\wsSj{ \wsSv{j} }
\def\wsSbi{ \wsSbv{i} }
\def\dwsS{ \del\wsS }

\def\wsSot{ \wsSv{12} }

\def\wsSpv#1{ \wsSp_{#1} }
\def\wsSpthf{ \wsSpv{34} }

\def\wsSpp{ \wsS^{\prime\prime} }
\def\wsSppv#1{ \wsSpp_{#1} }
\def\wsSppof{ \wsSppv{14} }

\def\wsfv#1#2{ (#1,#2) }
\def\wsfSG{ \wsfv{\wsS}{\xGa} }
\def\wsfSGp{ \wsfv{\wsSp}{\xGap} }

\def\wsfSGv#1{ \wsfv{\wsSv{#1}}{\xGav{#1}} }

\def\wsfSGj{ \wsfSGv{\xj} }
\def\xj{ j }

\def\wsSp{ \wsS\p }
\def\xGap{ \xGa\p }

\def\xn{ n }
\def\xm{ m }
\def\xW{ W }
\def\xWb{ \bar{\xW} }
\def\xe{ e }
\def\xP{ P }

\def\xT{ \mathcal{T} }
\def\xTb{ \bar{\xT} }
\def\xI{ \mathcal{I} }
\def\xM{ M }
\def\xD{ D }
\def\xG{ G }
\def\xF{ F }
\def\xv{ v }
\def\xH{ H }
\def\xA{ A }
\def\xC{ C }
\def\xO{ O }
\def\xd{ d }

\def\xt{ t }
\def\xg{ g }

\def\xN{ N }
\def\yyN{ N }
\def\yya{ a }

\def\yn{ n }

\def\xz{ z }
\def\xJ{ J }

\def\xi{ i }
\def\xj{ j }
\def\xxi{ k }
\def\xxj{ j }
\def\xxl{ l }
\def\yi{ i }
\def\yj{ j }
\def\yk{ k }

\def\yL{ L }

\def\xOv#1{ \xO_{#1} }
\def\xOo{ \xOv{1} }
\def\xOt{ \xOv{2} }

\def\xOsv#1{ \xO^{\ast}_{#1} }
\def\xOj{ \xOv{\xxj} }
\def\xOsj{ \xOsv{\xxj} }
\def\xOvr{ \xOv{\xv} }
\def\xOvrp{ \xOv{\xvp} }

\def\xmi{ \xm_{\xi} }
\def\xmj{ \xm_{j} }
\def\xWv#1{ \xW_{#1} }
\def\xWi{ \xWv{\xi} }
\def\xWm{ \xWv{\yym} }
\def\xWe{ \xWv{\xe} }
\def\xWye{ \xWv{\yed} }
\def\xWyvr{ \xWv{\yvr} }
\def\xWvr{ \xWv{\xv} }
\def\xWgm{ \xWv{\xgm} }
\def\xWa{ \xWv{\aplg} }
\def\xTv#1{ \xT_{#1} }
\def\xTi{ \xTv{\xi} }

\def\xTm{ \xTv{\yym} }

\def\xTbi{ \xTv{\xi} }

\def\xDv#1{ \xD_{#1} }
\def\xDvr{ \xDv{\xv} }
\def\xDgm{ \xDv{\xgm} }

\def\xHv#1{ \xH_{#1} }
\def\xHo{ \xHv{1} }
\def\xHt{ \xHv{2} }

\def\xHP{ \xHv{\xP} }

\def\xHvr{ \xHv{\xv} }
\def\xHvv#1#2{ \xH_{#1}^{#2} }
\def\xHvrv#1{ \xHvv{\xv}{#1} }
\def\xHvrz{ \xHvrv{0} }
\def\xHvro{ \xHvrv{1} }
\def\xHgmP{ \xHv{\xgmP} }

\def\xHgmo{ \xHv{\xgmo} }
\def\xHgmt{ \xHv{\xgmt} }

\def\xHgmst{ \xHv{\xgmst} }

\def\yI{ I }
\def\yIv#1{ \yI_{#1} }

\def\ICm{ \IC^{\xm} }
\def\ICmi{ \IC^{\xmi} }
\def\ICphi{ \IC[\bphi] }
\def\ICphii{ \IC[\bphi_\xi] }
\def\ICbphitot{ \IC[\bphitot] }

\def\bphitot{ \bphi_{\tot} }

\def\xGa{ \Gamma }

\def\xGav#1{ \xGa_{#1} }

\def\Gr{ \mathrm{Gr} }
\def\Grv#1#2{ \Gr_{#1,#2} }
\def\Grmin{ \Grv{\xm_i}{\xn} }

\def\xIv#1{ \xI_{#1} }
\def\xIe{ \xIv{\xe} }

\def\xIvr{ \xIv{\xv} }

\def\IHp{ \IH^+ }

\def\IHv#1#2{ \IH^{#1}_{#2} }
\def\IHpv#1{ \IHv{+}{#1} }

\def\IHpi{ \IHpv{\xi} }

\def\botiIvr{ \bigotimes\nolimits_{\xe\in\xIvr} }

\def\sIie{ \sum_{\xi\in\xIe} }

\def\prodiNSdet{ \prod_{\xi=1}^{\xNS}(\det\delj\delk\xWi)^{\xgSi} }

\def\mLgr{ L }
\def\mLgrv#1{ \mLgr_{#1} }

\def\mLgrTi{ \mLgrv{\xTi} }

\def\xTe{ \xTv{\xe} }
\def\mLgrTe{ \mLgrv{\xTe} }

\def\xTb{ \bar{\xT} }
\def\xTbi{ \xTb_{\xi} }

\def\thLGv#1#2{ \lrbc{#1;#2} }
\def\thLGpW{ \thLGv{\bphi}{\xW} }
\def\thLGpmW{ \thLGv{\bphi}{-\xW} }
\def\thLGpWi{ \thLGv{\bphii}{\xWi} }
\def\thLGpWm{ \thLGv{\bphim}{\xWm} }
\def\thLGepW{ \thLGv{\bphie}{\xWe} }

\def\thLGyeW{ \thLGv{\bphiye}{\xWye} }

\def\Zt{ \ZZ_2 }
\def\Ztgr{$\Zt$-graded}
\def\xMv#1{ \xM^{#1} }
\def\xMz{ \xMv{0} }
\def\xMo{ \xMv{1} }

\def\xMlv#1{ \xM_{#1} }

\def\xMvr{ \xMlv{\xv} }
\def\xMe{ \xMlv{\xe} }
\def\xMi{ \xMlv{\xxi} }
\def\xMes{ \xdu{\xMe} }
\def\xMgm{ \xMlv{\xgm} }
\def\xMa{ \xMlv{\aplg} }
\def\xMao{ \xMlv{\aplgo} }
\def\xMat{ \xMlv{\aplgt} }

\def\xMgms{ \xMlv{\xgms} }

\def\xMapst{ \xMlv{\aplgpst} }

\def\xMaso{ \xMlv{\aplgso} }
\def\xMast{ \xMlv{\aplgst} }

\def\xMapo{ \xMlv{\aplgpo} }
\def\xMappo{ \xMlv{\aplgppo} }
\def\xMapst{ \xMlv{\aplgpst} }
\def\xMappst{ \xMlv{\aplgppts} }

\def\xMlz{ \xMlv{0} }
\def\xMlo{ \xMlv{1} }
\def\xMlt{ \xMlv{2} }
\def\xMlzs{ \xdu{\xMlz} }

\def\xDv#1{ \xD_{#1} }
\def\xDz{ \xDv{0} }
\def\xDo{ \xDv{1} }

\def\xtMe{ \xtMv{\xe} }
\def\xtMe{ \xMe\otimes\xMes }

\def\xdu#1{ {#1}^{\ast} }
\def\xMs{ \xdu{\xM} }
\def\xDs{ \xdu{\xD} }
\def\xFs{ \xdu{\xF} }
\def\xGs{ \xdu{\xG} }
\def\xes{ \xdu{\xe} }

\def\Id{ \mathrm{Id} }

\def\mmfv#1#2#3{ \lrbc{#1,#2,#3} }
\def\mmfMDW{ \mmfv{\xM}{\xD}{\xW} }
\def\mmfMDWi#1{ \mmfv{\xM_{#1}}{\xD_{#1}}{\xW_{#1}} }
\def\mmfMDWe{ \mmfMDWi{\xe} }

\def\mmfMDWv{ \mmfMDWi{\xv} }
\def\mmfMDWl#1{ \mmfv{\xM_{#1}}{\xD_{#1}}{\xW} }
\def\mmfMDWxi{ \mmfMDWl{\xxi} }
\def\mmfMDWxj{ \mmfMDWl{\xxj} }
\def\mmfMDWz{ \mmfMDWl{0} }
\def\mmfMDWo{ \mmfMDWl{1} }

\def\mmfMDWxi{ \mmfMDWl{i} }
\def\mmfMDWxk{ \mmfMDWl{\xxi} }
\def\mmfMDWyed{ \mmfMDWi{\yvr} }
\def\mmfMDWgm{ \mmfMDWi{\xgm} }
\def\mmfMDWa{ \mmfMDWi{\aplg} }

\def\fcntr{ f_{\mathrm{cntr}} }
\def\fcntr{ f }
\def\ftot{ f_{\xGa} }

\def\bi{ \bar{\yi} }
\def\bj{ \bar{\yj} }
\def\xzb{ \bar{\xz} }

\def\etav#1{ \eta^{#1} }
\def\etaib{ \etav{\bi} }
\def\etajb{ \etav{\bj} }

\def\rhov#1#2{ \rho_{#1}^{#2} }
\def\rhozi{ \rhov{\xz}{\yi} }
\def\rhozbi{ \rhov{\xzb}{\yi} }
\def\rhozbj{ \rhov{\xzb}{\yj} }

\def\thetav#1{ \theta_{#1} }
\def\thetai{ \thetav{\yi} }

\def\Lgv#1{ \yL_{#1} }
\def\LgT{ \Lgv{\xT} }
\def\LgTb{ \Lgv{\xTb} }
\def\delv#1{ \del_{#1} }
\def\delz{ \delv{\xz} }
\def\delzb{ \delv{\xzb} }
\def\deli{ \delv{\yi} }
\def\delj{ \delv{\yj} }
\def\delk{ \delv{\yk} }
\def\delib{ \delv{\bi} }
\def\deljb{ \delv{\bj} }

\def\phiv#1{ \phi^{#1} }
\def\phii{ \phiv{\yi} }
\def\philv#1{ \phi_{#1} }
\def\philvv#1#2{ \philv{#1,#2} }

\def\philij{ \philvv{\yym}{\yyj} }

\def\philmio{ \philvv{\yymi}{1} }

\def\philmo{ \philvv{\yym}{1} }
\def\phiib{ \phiv{\bi} }

\def\Qbrst{ Q }
\def\vQ{ \delta_\Qbrst }

\def\intv#1{ \int_{#1} }
\def\intwsS{ \intv{\wsS} }
\def\intdwsS{ \intv{\dwsS} }

\def\ointCi{ \oint_{\xCi} }

\def\rhodf{ \rhozi dz + \rhozbi d\zb }
\def\brhodf{ (\rhodf) }
\def\xWId{ \xW\,\Id }

\def\xCv#1{ \xC_{#1} }
\def\xCi{ \xCv{\xxi} }
\def\ined{ \xxi,\xxl }
\def\xekl{ \xev{\ined} }

\def\xOkl{ \xOv{\ined} }

\def\STrMlv#1{ \STr_{\xMlv{#1}} }

\def\xcAv#1{ \xA_{#1} }
\def\xAT{ \xcAv{\xT} }
\def\xAe{ \xcAv{\xe} }
\def\xATe{ \xcAv{\xTe} }

\def\corrSG#1{ \corri{#1}{\wsfSG} }
\def\corrSGp#1{ \corri{#1}{\wsfSGp} }

\def\corrOPOvSG{ \corrSG{
\pxcPOP
\pxcVOv } }
\def\corrIOPOvSG{ \corrSG{\yIiot\pxcPpOP
\pxcVpOv}}
\def\corrOPOvSGp{ \corrSGp{\pxcPpOP
\pxcVpOv}}

\def\tot{ \wsS }

\def\HmMoz{ \Hom(\xMlz,\xMlo) }
\def\xMlozs{ \xMlo\otimes\xMlzs }

\def\Jv#1{ \xJ_{#1} }
\def\JW{ \Jv{\xW} }
\def\JWi{ \Jv{\xWi} }

\def\xOP{ \xOv{\xP} }

\def\xPv#1{ \xP_{#1} }

\def\xPo{ \xPv{1} }
\def\xPt{ \xPv{2} }

\def\xPth{ \xPv{3} }
\def\xPf{ \xPv{4} }

\def\tpii{ 2\pi i }

\def\Trv#1{ \Tr_{#1} }
\def\TrW{ \Trv{\xW} }

\def\St{ S^2 }

\def\dijW{ \det\deli\delj\xW }
\def\iddW{ \bdel \xW }

\def\xev#1{ \xe_{#1} }
\def\xez{ \xev{0} }
\def\xDe{ \xDv{\xe} }
\def\xDez{ \xDv{\xez} }
\def\dlvw#1{ \del#1^{\wedge} }
\def\dlDw{ \dlvw{\xD} }
\def\xDvv#1#2{ \xDv{#1,#2} }
\def\dlDkow{ \dlvw{\xDvv{\xxi}{1} } }
\def\delsv#1{ \delv{\sigma(#1)} }

\def\xxn{ n }
\def\xxnv#1{ \xxn_{#1} }
\def\xxnk{ \xxnv{\xxi} }

\def\rS{ \mathrm{S} }
\def\rSv#1{ \rS_{#1} }
\def\rSm{ \rSv{\xm} }
\def\rSmi{ \rSv{\xmi} }

\def\xOCi{ \xOv{\xCi} }
\def\xOGa{ \xOv{\xGa} }

\def\xNv#1{ \xN_{#1} }
\def\xNS{ \xNv{\wsS} }
\def\bphiS{ \bphiv{\wsS} }
\def\ICbphiS{ \IC[\bphiS] }
\def\TrS{ \Trv{\wsS} }

\def\xgv#1{ \xg(#1) }
\def\xgSi{ \xgv{\wsSi} }

\def\Wil{ \mathcal{W} }
\def\Wilv#1{ \Wil_{#1} }
\def\WilCk{ \Wilv{\xCi} }
\def\WilC{ \Wilv{\xC} }
\def\WilGa{ \Wilv{\xGa} }

\def\xij{ \xi,\xj }
\def\Cij{ \xCv{\xij} }
\def\eij{ \xev{\xij} }
\def\dlDij{ \dlvw{\xDv{\xij} } }
\def\dlDi{ \dlvw{\xDv{i} } }
\def\dlDj{ \dlvw{\xDv{j} } }
\def\xMij{ \xMlv{\eij} }

\def\xOe{ \xOv{\xe} }

\def\yN{ \xgm }
\def\yNv#1{ \yN_{#1} }
\def\yNP{ \yNv{\xP} }

\def\xPp{ \xP\p }

\def\xB{ S }
\def\xBv#1{ \xB_{(#1)} }
\def\xBvv#1#2{ \xBv{#1,#2} }

\def\xBot{ \xBvv{1}{2} }

\def\xU{ U }
\def\xUv#1{ \xU_{#1} }
\def\xUP{ \xUv{\xP} }

\def\xUo{ \xUv{1} }
\def\xUt{ \xUv{2} }

\def\ynod{$(\yn+1)$-dimensional}
\def\ynod{2-dimensional}

\def\dprv#1#2#3{ (#1,#2) }
\def\dprPpv#1#2{ \dprv{#1}{#2}{\xP,\xPp} }

\def\yI{ I }
\def\yIv#1{ \yI_{#1} }
\def\yIvv#1#2{ \yIv{#1,#2} }
\def\yIot{ \yIvv{1}{2} }

\def\yIgsg{ \yIvv{\xgm}{\xgms} }
\def\yIiot{ \yIivv{1}{2} }
\def\yIiv#1{ \yI^{-1}_{#1} }
\def\yIivv#1#2{ \yIiv{#1,#2} }
\def\yIgmts{ \yIvv{\xgmst}{\xgmt} }

\def\xgm{ \gamma }

\def\xgms{ \xdu{\xgm} }
\def\xgmv#1{ \xgm_{#1}^{} }

\def\xgmsv#1{ \xgm_{#1}^{\ast} }
\def\xgmsf{ \xgmsv{4} }

\def\xgmP{ \xgmv{\xP} }
\def\xgmj{ \xgmv{\xj} }

\def\xgmst{ \xgmsv{2} }

\def\xHgm{ \xHv{\xgm} }

\def\xHgms{ \xHv{\xgms} }

\def\xgmo{ \xgmv{1} }
\def\xgmt{ \xgmv{2} }
\def\xgmth{ \xgmv{3} }

\def\xgmos{ \xgmsv{1} }

\def\Cnv#1{ \mathrm{C}#1 }
\def\Cngm{ \Cnv{\xgm} }
\def\CngmP{ \Cnv{\xgmP} }

\def\Suspv#1{ \Sigma#1 }
\def\Suspgm{ \Suspv{\xgm} }

\def\xvrv#1{ \xv_{#1} }
\def\xvro{ \xvrv{1} }
\def\xvrt{ \xvrv{2} }

\def\xvrj{ \xvrv{\xj} }

\def\xvp{ \xv\p }

\def\yvr{ \nu }

\def\xOp{ \xO\p }
\def\xIgs{ \yIivv{\xgm}{\xgms} }

\def\silg{ \Upsilon }
\def\silgv#1{ \silg_{#1} }
\def\silgyvr{ \silgv{\yvr} }

\def\Lv#1{ \cL_{#1} }

\def\La{ \Lv{\aplg} }

\def\yyj{ j }
\def\yyc{ c }
\def\yycv#1{ \yyc_{#1} }

\def\yycmi{ \yycv{\yymi} }

\def\yycvv#1#2#3{ \yycv{#1}(#2;#3) }
\def\yyciphit{ \yycvv{\yym}{\bphim}{\yyt} }
\def\yycsv#1#2{ \yycv{#1,#2} }

\def\yycmj{ \yycsv{\yym}{\yyj} }

\def\yycmNo{ \yycsv{\yym}{\yyN+1} }

\def\yycmjphim{ \yycmj(\bphim) }

\def\yyt{ t }

\def\ICm{ \ICv{\yym} }

\def\GrmN{ \Grv{\yym}{\yyN} }

\def\GrmiN{ \Grv{\yymi}{\yyN} }
\def\yyp{ p }
\def\yypv#1{ \yyp_{#1} }
\def\yypj{ \yypv{\yyj} }
\def\yypo{ \yypv{1} }
\def\byyp{ \mathbf{\yyp} }
\def\tbyyp{ \tilde{\byyp} }
\def\yyq{ q }
\def\yyqv#1{ \yyq_{#1} }
\def\yyqj{ \yyqv{\yyj} }
\def\byyq{ \mathbf{\yyq} }
\def\Kmf#1#2{ (#1;#2) }
\def\yyn{ n }

\def\yymi{ \yym_\yyi }

\def\xWbm{ \xWv{\yybm} }

\def\xWmi{ \xWv{\yymi} }

\def\syyi{ \sum_{\yyi} }

\def\syyjoi{ \sum_{\yyj=1}^\infty }
\def\sjoNot{ \sum_{1\leq\yyj\leq{\yyN+1\over 2} } }

\def\xWN{ \xWv{\yyN} }

\def\xlist#1#2{ (#1\,|\,#2) }
\def\yyr{ r }
\def\yyrv#1{ \yyr_{#1} }
\def\yyrj{ \yyrv{\yyj} }
\def\SUN{ {\rm SU}(\yyN) }
\def\yyV{ V }

\def\yym{ m }
\def\yyi{ i }
\def\yybm{ \mathbf{\yym} }

\def\xcP{ \mathcal{P} }
\def\pxcP{ \prod_{\xP\in\xcP} }
\def\pxcPOP{ \pxcP\xOP }
\def\pxcPOPbphi{ \pxcPOP(\bphi) }

\def\xcPp{ \xcP\p }
\def\pxcPp{ \prod_{\xP\in\xcPp} }
\def\pxcPpOP{ \pxcPp\xOP }

\def\xA{ A }
\def\xAv#1{ \xA_{#1} }

\def\xAaot{ \xAv{\aplgo,\aplgst} }
\def\xAao{ \xAv{\aplgo,\aplgpso} }
\def\xAat{ \xAv{\aplgt,\aplgpst} }
\def\xAapot{ \xAv{\aplgpo,\aplgpts} }
\def\xAappot{ \xAv{\aplgppo,\aplgppts} }

\def\xR{ R }
\def\xRv#1{ \xR_{#1} }

\def\xRa{ \xRv{\aplg} }
\def\otv#1{ \otimes_{#1} }
\def\otR{ \otv{\xR} }
\def\Homv#1{ \Hom_{#1} }
\def\HomR{ \Homv{\xR} }

\def\yfv#1{ f_{#1} }
\def\yfot{ \yfv{12} }
\def\yfto{ \yfv{21} }

\def\Endv#1{ \End_{#1} }
\def\EndR{ \Endv{\xR} }

\def\xcV{ \mathcal{V} }
\def\xcVp{ \xcV\p }
\def\xcE{ \mathcal{E} }

\def\botvV{ \bigotimes_{\xv\in\xcV} }
\def\boteE{ \bigotimes_{\xe\in\xcE} }
\def\pxcV{ \prod_{\xv\in\xcV} }
\def\pxcVOv{ \pxcV\xOvr }

\def\pxcVp{ \prod_{\xv\in\xcVp} }
\def\pxcVpp{ \prod_{\xvp\in\xcVp} }
\def\pxcVpOv{ \pxcVp\xOvr }
\def\pxcVpOvp{ \pxcVpp\xOvrp }

\def\aplg{ \alpha }

\def\aplgv#1{ \aplg_{#1}^{} }

\def\aplgo{ \aplgv{1} }
\def\aplgt{ \aplgv{2} }
\def\aplgj{ \aplgv{\xj} }
\def\aplgp{ \aplg\p }
\def\aplgpv#1{ \aplgp_{#1} }

\def\aplgpp{ \aplg^{\prime\prime} }
\def\aplgppv#1{ \aplgpp_{#1} }
\def\aplgppj{ \aplgppv{\xj} }
\def\aplgppo{ \aplgppv{1} }
\def\aplgppt{ \aplgppv{2} }
\def\aplgppts{ \xdu{(\aplgppt)} }
\def\aplgs{ \xdu{\aplg} }
\def\aplgsv#1{ \aplgs_{#1} }
\def\aplgso{ \aplgsv{1} }
\def\aplgst{ \aplgsv{2} }

\def\aplgpsv#1{ \xdu{(\aplgpv{#1})} }
\def\aplgpso{ \aplgpsv{1} }
\def\aplgpst{ \aplgpsv{2} }

\def\aplgpv#1{ \aplgp_{#1} }
\def\aplgpj{ \aplgpv{\xj} }
\def\aplgpo{ \aplgpv{1} }
\def\aplgpt{ \aplgpv{2} }
\def\aplgpts{ \xdu{(\aplgpt)} }

\def\xAvv#1#2{ \xA[#1,#2] }
\def\xAgost{ \xAvv{\xgmo}{\xgmst} }
\def\xAgthsf{ \xAvv{\xgmth}{\xgmsf} }
\def\xAgosf{ \xAvv{\xgmo}{\xgmsf} }
\def\xAv#1{ \xA[#1] }

\def\xAgm{ \xAv{\xgm} }

\def\xncwsfv#1#2#3{ \wsfv{#1}{#2}_{#3} }
\def\xncwsfSGvv#1#2{ \xncwsfv{\wsS_{#1}}{\xGa_{#1}}{#2} }
\def\xncwsfSGvrv#1{ \xncwsfSGvv{#1}{\xv_{#1} } }
\def\xncwsfSGvro{ \xncwsfSGvrv{1} }
\def\xncwsfSGvrt{ \xncwsfSGvrv{2} }
\def\xncwsfSGv#1{ \xncwsfv{\wsS}{\xGa}{#1} }
\def\xncwsfSGPot{ \xncwsfSGv{12} }
\def\xncwsfSGPvr{ \xncwsfSGv{\xv} }
\def\xncwsfSGPvot{ \xncwsfSGv{\xvro,\xvrt} }
\def\xspvotv#1{ \wsfSG^{#1}_{\xvro,\xvrt} }
\def\xspvotp{ \xspvotv{\prime} }
\def\xspvotpp{ \xspvotv{\prime\prime} }

\def\xp{ p }
\def\xpvv#1#2{ \xp_{#1,#2} }
\def\xpvk#1{ \xpvv{#1}{\xk} }
\def\xpjk{ \xpvk{\xj} }
\def\xk{ k }
\def\xpok{ \xpvk{1} }
\def\xptk{ \xpvk{2} }

\def\xik{ \xi(\xk) }
\def\wsSik{ \wsSv{\xik} }

\def\cac{ c }
\def\cacv#1{ \cac_{#1} }
\def\cack{ \cacv{\xk} }

\def\dlfvv#1#2{ \del_{\phi_{#1,#2}^{} } }
\def\dlfik{ \dlfvv{i}{k} }
\def\dlfjl{ \dlfvv{j}{l} }
\def\dlfikDe{ \dlfik\xDe }
\def\dlfjlDe{ \dlfjl\xDe }

\newcommand{\bimnx}[7]{\bibitem{#1}#2,
{\em #3}, { #4}\hspace{0.25em}{\bf
#5}\hspace{0.25em}(#6)\hspace{0.25em}{#7}}

\begin{document}

\setlength{\unitlength}{3947sp}
\setlength{\unitlength}{1mm}

\begin{titlepage}
\vfill
\begin{center}

{\large \bf
Topological \LG\ models on a \wsf}\\

\bigskip

\bigskip
\centerline{\textsc{M. Khovanov\footnotemark[1]}}

\centerline{\em Department of Mathematics, University of
California, Davis}
\centerline{\em 1 Shields Ave.}
\centerline{\em Davis, CA 95616-8633}
\centerline{{\em E-mail address:} {\tt mikhail@math.ucdavis.edu}}

\bigskip
\centerline{\textsc{L. Rozansky}\footnote[1]{ This work was supported by NSF
Grants DMS-0104139 and DMS-0196131} }

\centerline{\em Department of Mathematics, University of North
Carolina} \centerline{\em CB \#3250, Phillips Hall}
\centerline{\em Chapel Hill, NC 27599} \centerline{{\em E-mail
address:} {\tt rozansky@math.unc.edu}}

\vfill
{\bf Abstract}

\end{center}
\begin{quotation}

We define \tLGms\ on a \wsf, that is, on a collection of
2-dimensional surfaces whose boundaries are sewn together along
the edges of a graph. We use matrix factorizations in order to
formulate the boundary conditions at these edges and produce
a formula for the \crrs. Finally, we present the \gluing\
formulas, which correspond to various ways in
which the pieces of a \wsf\ can be joined together.

\end{quotation} \vfill \end{titlepage}

\pagebreak
\tableofcontents
\section{Introduction}

It is always easier to define a \qft\ on a closed manifold:
there is no need to formulate the boundary conditions for the
fields in the path integral. If the boundary exists, then one
might limit oneself to the easiest case of the Neumann boundary conditions.
The last decade showed, however, that
the world of boundary conditions may be even more interesting and
diverse than the world of the `\bulk' \QFTs. A \bulk\ 2-dimensional
\QFT\ yields an algebra, boundary conditions are objects of a category, and all morphisms between
two objects form a module over the \bulk\ algebra.
Now it turns out that the existence of
a rather general class of boundary conditions may change the very
nature of the \ws\ manifold: instead of being just a surface with
boundary, it may become a `\foam', that is, a version of a
2-dimensional \CW-complex endowed with a complex structure, if
needed.

Although the \QFTs\ themselves do not require the presence of a \wsf,
a \foam\ appeared in the paper\cx{Kh1} as a
necessary element in the categorification of the $\SU(3)$ \HOMFLY\
polynomial. The paper\cx{Ro1} interpreted that part of the
categorification as a \tdm\ topological \Amdl\ defined on a \wsf.
Each 2-dimensional connected component $\wsSi$ of the \foam\ carries its own topological
\smdl, whose target space is the complex \Grmn\ $\Grmin$, while the
boundary condition at an edge of the \smgr\ is specified by
selecting a \lgsm\ in the cross-product of the \Grmns\
assigned to the components $\wsSi$ bounding the edge.

The paper\cx{Ro1} described the general setup of a \QFT\ on a
\wsf, using \tAmdls\ as an illustration.
\TAms, however,
are notorious for their complexity even on the usual smooth surfaces, and
the paper\cx{Ro1} presented neither an accurate description of the
\Hlbss\ corresponding to the \smgr\ vertices, nor a complete
formula for the \pf.

In this paper we give a detailed
description of a \tLGm\ on a \wsf. Each component $\wsSi$ of the
\foam\ carries its own fields $\bphii=
\philvv{i}{1},\ldots,\philvv{i}{\xmi}$
and its
own potential $\xWi(\bphii)$, while each edge of the \sgr\
carries a matrix factorization of the sum of potentials of the
bounding components $\wsSi$, in the spirit of\cx{KL2}.
We describe the \Hlbss\ of the vertices of the \sgr\ and also provide
a formula for the \crrs, which generalizes the formulas of Vafa\cx{V}
and Kapustin-Li\cx{KL2}. Finally, we present the \gluing\ formulas
for joining `space-like' and even `time-like' boundary components of the \wsf.

This paper is closely related to our categorification\cx{KR} of the $\SU (N)$
\HOMFLY\
polynomial. Although we do not use \wsfs\ explicitly in\cx{KR}, the
construction of the graded vector spaces associated to 3-valent
graphs in\cx{KR} is a particular case of the definition of an
operator space $\xHgm$ related to a \dlcgr\ $\xgm$, as described
in subsection\rw{s5.2}. We refer the reader to\cx{KR} for a
detailed discussion of \mfs.

\section{A \xTLGT\ on a \ws\ with a boundary}

According to\cx{Ro1}, any \QFT\ defined on a 2-dimensional surface
with boundary can be transferred to a \wsf. Hence we begin by
reviewing the \xTLGT\ $\xT$ on a surface with boundary, as presented in\cx{La1}
(see also references therein). We assume for simplicity that the
target space of $\xT$ is a flat $\ICm$ and there are no gauge
fields. Then the \bulk\ theory is characterized by a (polynomial) \sp\ $\xW\in\ICphi$,
$\bphi=\seqllv{\phi}{\xm}$, and we denote this \LGt\ by $\xT=\thLGpW$.

\subsection{The \bLgr}

The \xTLGT\ $\xT$ contains bosonic fields $\bphi$ and $\bbphi$ as
well as the fermionic fields $\etaib$, $\thetai$, $\rhozi$ and
$\rhozbi$. The \bLgr\ of the theory is
\qq
\begin{split}
\LgT & =  \hlfv\lrbc{\delz\phii\,\delzb\phiib + \delz \phiib\,\delzb\phii
- \rhozi\,\delzb\etaib -\rhozbi\,\delz\etaib}
\\
&\quad
- 2i\Big(
\thetai\lrbc{\delz\rhozbi - \delzb\rhozi} + \deli\delj
\xW\,\rhozi\rhozbj
\Big)
\\
&\quad
+{1\over 4}\lrbc{
\delib\deljb\xWb\,\thetai\etajb - \deli\xW\,\delib\xWb
}.
\end{split}
\label{1.1}
\qqq
Each line in this expression is invariant under the topological
\BRST\ transformation $\Qbrst$
\begin{align}
\vQ\phii & = 0 & \vQ\phiib &= \etaib
\nonumber
\\
\vQ\etaib & = 0 & \vQ\thetai & = \deli\xW
\label{1.2}
\\
\vQ\rhozi & =  \delz\phii & \vQ\rhozbi &= \delib\phii,
\nonumber
\end{align}
except that the second line generates the boundary Warner term:
for a \ws\ $\wsS$ with a boundary $\dwsS$ the \BRST\ variation
of the action is
\qq
\vQ\intwsS \LgT = \intdwsS \deli\xW\brhodf.
\label{1.3}
\qqq

Note that if we treat $(\rhozi,\rhozbi)$ as a 1-form on $\wsS$,
then the \Lgr\rx{1.1} can be written without a reference to the
complex structure of the \ws. The only remnant of that complex
structure would be the orientation that it induces on $\wsS$.
Let $\LgTb$ denote the \Lgr\rx{1.1} in which we conjugated the
complex structure or, equivalently, reversed the orientation of
the \ws. It is easy to see that this change can be compensated by
switching two signs: the sign of the field $\thetai$ (that is, $\thetai$ is a `pseudo-scalar')
and the sign of the
\sp\ $\xW$. Thus,
\qq
\Lgv{\overline{\thLGpW}} = \Lgv{\thLGpmW}.
\label{1.4}
\qqq
\subsection{The boundary \Wl}
M.~Kontsevich suggested that the Warner term\rx{1.3} could be
compensated by putting the appropriate \Wls\ at the boundary
components of $\wsS$. This idea was implemented in
papers\cx{KL2},\cx{BHLS} and\cx{La1}. We will follow the approach
of Lazaroiu\cx{La1} as the most suitable for our purposes.

The linear space for a \aLG\ \Wl\ is provided by a \mf\ of the \sp\ $\xW$.
According
to\cx{KL2}, a \emph{\mf} of $\xW$ is a \trpl\ $\mmfMDW$, where $\xM$ is a finite-dimensional \Ztgr\
free $\ICphi$
module, $\xM = \xMz \oplus \xMo$ ($\rank \xMz = \rank \xMo$), while the \emph{\twdif} $\xD$
is an operator $\xD\in\End(\xM)$, such that $\deg \xD=1$ and
\qq
\xD^2 = \xW\,\Id.
\label{1.4x}
\qqq
Simply saying, a \mf\ is described by a $2n\times
2n$-dimensional matrix with polynomial entries
\begin{eqnarray}
\xD = \lrbc{
\begin{matrix}
0 & \multicolumn{1}{|c}{\xF}
\\
\cline{1-2}
\xG &\multicolumn{1}{|c}{0}
\end{matrix}
},\;\;\mbox{such that}\quad \xF\xG = \xG\xF = \xWId.
\label{1.5}
\end{eqnarray}
Lazaroiu introduces a \sc
\qq
\xAT =
\lrbc{
\begin{matrix}
\xWId & \multicolumn{1}{|c}{\brhodf\,\deli\xF}
\\
\cline{1-2}
\brhodf\,\deli\xG &\multicolumn{1}{|c}{\xWId}
\end{matrix}
}
\label{1.6}
\qqq
acting on $\xM$. Suppose that $\dwsS$ is a union of disjoint
circles:
\qq
\dwsS=\bigsqcup_\xxi \xCi.
\label{1.6x}
\qqq
To each circle $\xCi$ we
assign a \mf\ $\mmfMDWxk$ of $\xW$. Then Lazaroiu proves
that the \pigr\
\qq
\exp\lrbc{\intwsS \LgT}\,\prod_\xxi \WilCk,\quad\WilCk = \STr_{\xMi} \Pexp \ointCi \xA,
\label{1.7}
\qqq
with the orientation of
$\xCi$ induced by the orientation of $\wsS$,
is invariant under the topological \BRST\ transformation\rx{1.2}.

An orientation of a boundary component $\xCi$ can be reversed
without affecting its \st, if we replace the associated \mf\
$\mmfMDWxk$ with its \dual. First, observe that the \mfs\ can
be tensored:
\qq
\mmfMDWi{1}\otimes\mmfMDWi{2} =
\mmfv{\xM_1\otimes\xM_2}{\xD_1+\xD_2}{\xW_1+\xW_2}.
\label{1.9}
\qqq
Then we define the dual \mf\ as
\qq
\xdu{\mmfMDW} =
\mmfv{\xMs}{\xDs}{-\xW},
\label{1.10}
\qqq
where the module $\xMs$ is the dual of
$\xM$ over $\ICphi$ and
\begin{eqnarray}
\xDs = \lrbc{
\begin{matrix}
0 & \multicolumn{1}{|c}{\xGs}
\\
\cline{1-2}
-\xFs &\multicolumn{1}{|c}{0}
\end{matrix}
},
\label{1.11}
\end{eqnarray}
where $\xFs$ and $\xGs$ are the dual maps (transposed matrices) of $\xF$ and
$\xG$. The choice\rx{1.11} guarantees that the natural pairing map
$\xMs\otimes\xM\xrightarrow{\fcntr}\ICphi$ satisfies the property
$\fcntr\xD = 0$ and thus `commutes' with $\xD$.

\subsection{Boundary operators}
In order to simplify our notations, whenever we work with multiple
\mfs, we will denote all their \twdifs\ by $\xD$, if it is clear
on which particular module that $\xD$ is acting. Also, let
$\scommm{\cdot}{\cdot}$ denote the \scommu:
\qq
\scommm{A}{B} = AB - (-1)^{\deg A\,\deg B}\, BA.
\label{sc}
\qqq

The \sts\ of\rx{1.7} have an obvious generalization. Let
$\xt\in[0,1]$ parameterize a boundary circle $\xC$. The values $0=\xt_0\leq
\xt_1\leq\cdots\leq \xt_\yn=1$ split $\xC$ into $\yn$ segments. We
can assign any \mfs\ $\mmfMDWxj$ to the segments
$[\xt_{\xxj-1},\xt_\xxj]$. To each value $\xt_j$ we assign an
operator $\xOj\in\Hom(\xMlv{j},\xMlv{j+1})$, which \scoms\ with
the \twdif\
$\xD$:
$\scommm{\xD}{\xOj}=0$.
Then
the \Wl\ contribution $\WilC$
can be replaced in the integrand\rx{1.7} by
\qq
\WilC = \STrMlv{1}\lrbc{ \Pexp\int_{[\xt_0,\xt_1]}\xA}\xOv{1}\cdots
\lrbc{ \Pexp\int_{[\xt_{\yn-1},\xt_{\yn}]}\xA}\xOv{\yn},
\label{1.8}
\qqq
while still maintaining the \BRST\ invariance.
%

Following \cx{KL2}, let us give a more precise description of the space of the
operators $\xOj$. For two \mfs\ $\mmfMDWz$ and
$\mmfMDWo$ consider the \Ztgr\ module $\HmMoz$
of $\ICphi$-linear
maps between the modules. Define a differential $\xd$ on this module by
%
\qq
\xd\xO = \scommm{\xD}{\xO},
\label{1.8y1}
\qqq
where $\xO\in\HmMoz$.
It turns out that $\xd^2=0$, and $\xd$ describes the \BRST\ action on
$\HmMoz$.
The space $\xHP$ of operators at the junction $\xP$ of two segments
carrying the modules $\xMlz$ and $\xMlo$ can be presented as
%
\qq
\xHP = \Ext(\xMlz,\xMlo) = \ker\xd/\im\xd.
\label{1.8x2}
\qqq
%

An equivalent presentation of the operator space comes from the dual
\mf. Namely, consider the tensor product of \mfs
\qq
\mmfMDWo\otimes\xdu{\mmfMDWz} = \mmfv{\xMlozs}{\xD}{0},\qquad
\xD = \xDo + \xdu{\xDz}.
\label{1.8x1}
\qqq
Since $\xD^2=0$, we can take
\qq
\xHP=\ker\xD/\im\xD,
\label{1.8x3}
\qqq
because the
cohomology of $\xD$ is
canonically isomorphic to $\Ext(\xMlo,\xMlz)$ in view of the
canonical isomorphism
\qq
\HmMoz = \xMlo\otimes\xMlzs,
\label{1.8x}
\qqq
and the fact that $\xd$ corresponds to $\xD$.

\section{A \xTLGT\ on a \wsf}
\subsection{The \wsf}
Let us recall the definition of a \wsf\ given in\cx{Ro1}. Let
$\xGa$ be a graph such that every vertex has \adjacent\ edges.
$\xGa$ is allowed to contain disjoint circles. A \ccl\ on $\xGa$
is defined to be either a disjoint circle or a cyclicly ordered finite sequence of edges, such
that the beginning of the next edge corresponds to the end of the
previous edge. Let $\wsS$ be an orientable and possibly disconnected smooth 2-dimensional surface, its
boundary $\dwsS$ being a union of disjoint circles. A \emph{\wsf}
$\wsfSG$ is a union $\xGa\cup\wsS$, in which the boundary circles
of $\wsS$ are glued to some cycles on $\xGa$ in such a way that
every edge of $\xGa$ is glued to at least one circle of $\dwsS$.

Defining a \xTLGT\ on a \wsf\ $\wsfSG$ involves three steps:
first, we assign \bulk\ theories to oriented connected components of
$\wsS$; second, we assign appropriate boundary condition to
the oriented \smeds, and third, we choose the operators at the \smvrs. The
first two steps comprise a \emph{\dcrtn} of the \wsf.
%

\subsection{Bulk theories on the 2-dimensional connected components}
The first step is simple. Suppose that the 2-dimensional surface
$\wsS$ is a union of $\xNS$ disjoint components $\wsS=\bigsqcup_{i=1}^{\xNS}\wsSi$.
Then to every oriented component $\wsSi$ we assign its own \xTLGT\
$\xTi=\thLGpWi$ with its own \ts\ $\ICmi$, bosonic fields $\bphii$,
fermionic fields and a \sp\ $\xWi$ in such a way that if $\wsSi$
and $\wsSbi$ represent the same component of $\wsS$ with opposite
orientations, then $\wsSbi$ should be assigned the \conjugated\
theory $\xTbi$.

\subsection{Boundary conditions at the \sgr\ edges}

Next, we have to formulate the boundary conditions at the \smeds,
which would be compatible with the \BRSTinv\ of the actions of \LG\
theories sitting on the adjacent surfaces $\wsSi$. According
to\cx{Ro1}, these boundary conditions are just particular cases of an
ordinary boundary condition for a \LGt\ defined on a smooth
surface with a boundary.

Let us orient an edge $\xe$ of $\xGa$ and let $\xIe$ be the set of
indices $\xi$, such that $\xe$ bounds $\wsSi$. We orient all
$\wsSi$ ($\xi\in\xIe$) in such a way that their orientations are
compatible with the orientation of $\xe$. If $\xP$ is an inner
point of $\xe$, then a small neighborhood of $\xP$ looks like a
union of upper half-planes $\IHpi\subset\IC$ glued along the common
real line, the point $\xP$ being the origin of that real line. If
$\IHp$ is a `standard' upper half-plane, then we can identify all
$\IHpi$ analytically (preserving orientation) with it.
Thus, if every $\IHpi$ carries a \QFT\ $\xTi$ with the \xLgr\
$\mLgrTi$, then formulating a boundary condition for them at
$\xe$ is equivalent to formulating it for the combined theory
$\xTe$ with the \xLgr\ $\mLgrTe = \sum_{\xi\in\xIe}\mLgrTi$. In
case of the \xTLGTs\ this means that to every \smed\ $\xe$ we
assign the theory
\qq
\xTe = \thLGepW,\;\;\mbox{where}\qquad\bphie = (\bphi\,|\,\xi\in\xIe),\qquad
\xWe = \sIie\xWi.
\label{2.1}
\qqq
Then to every oriented edge $\xe$ of the \sgr\ $\xGa$ we associate a
\mf\ $\mmfMDWe$ in such a way that if two oriented edges $\xe$
and $\xes$ represent the same edge with opposite orientations,
then $\mmfMDWi{\xes} = \xdu{\mmfMDWe}$. Also, we assign to $\xe$
the Lazaroiu connection
$\xAe=\xATe$.
If the \mf\ $\mmfMDWe$ does
not factor into a tensor product of \mfs\ of individual \sps\
$\xWi$, then the boundaries of the components $\wsSi$ can not be
`unglued' at the edge $\xe$.

It is important to note that the construction of a \mf\ associated
to a \smed\ must be local. It may happen that because of the global structure
of the \wsf\ $\wsfSG$, some of the strips
that bound an edge $\xe$ come from the same \ws\ component
$\wsSi$. In this case, we first treat their theories $\xTi$ as
different, that is, they share the same dimension of the target
space $\xmi$ and the same \sp\ $\xWi$, yet their target spaces and
fields are considered distinct. After we pick a \mf\ $\mmfMDWe$,
we impose a condition that the fields coming from the different
strips of the same \ws\ component are the same.

\subsection{Operators at the \sgr\ vertices}

Let $\xv$ be a \smgrvr\ and let $\xIvr$ be the set of \smeds\ \adjacent\ to $\xv$.
We orient these edges away from $\xv$ and consider the
factorization
\qq
\mmfMDWv = \botiIvr
\mmfMDWe.
\label{2.5}
\qqq
Obviously,
\qq
\xMvr = \botiIvr\xMe, \qquad
\xWvr =0,
\label{2.6}
\qqq
the latter equation follows, since for every
component $\wsSi$ attached to $\xv$ there are two (or, more
generally, an even number of) bounding edges, which are attached
to $\xv$ in such a way that $\wsSi$ contributes an equal number of $\xWi$ and
$-\xWi$ to $\xWvr$.

Since
$\xWvr=0$, then $\xDvr^2=0$ and we can consider its cohomology
\qq
\xHvr = \ker \xDvr / \im \xDvr.
\label{2.7}
\qqq
The space $\xHvr$ is \Ztgr: $\xHvr =
\xHvrz\oplus\xHvro$. $\xDvr$ plays the role of the \BRST\ operator at $\xv$, so we take $\xHvr$ as the space of operators at the vertex $\xv$. In other words, to
every vertex $\xv$ we associate an element
\qq
\xOvr\in\ker\xDvr,
\label{2.7x}
\qqq
and the \BRSTinv\ of the path integral will guarantee that
the correlators depend on $\xOvr$ only modulo $\im \xDvr$.

\subsection{\Wntw}
For a \wsf\ $\wsfSG$, the analog of the \Wls\ at the boundary
components of the \ws\ $\wsS$ is the \Wntw\ formed by the \sgr\
$\xGa$. Its contribution is a generalization of the \st\ \rx{1.8} and
it is expressed through multiple
contractions between the pairs of dual modules in a big tensor
product
\qq
\lrbc{
\botvV
\xMvr } \otimes
\lrbc{
\boteE
(\xtMe)},
\label{2.8}
\qqq
where $\xcV$ and $\xcE$ are the sets of all vertices and of all
edges of $\xGa$.
If we substitute the formulas\rx{2.6} for $\xMvr$, then all the
elementary modules $\xMe$ and $\xMes$ can be grouped in pairs of
mutually \dual\ modules: $\xMe$ (or $\xMes$) coming from
$\xtMe$ and the \dual\ module $\xMes$ (or $\xMe$) coming from
the tensor product expression\rx{2.6} for $\xMvr$, where $\xv$ is the beginning (or the
end point) of the oriented edge $\xe$. Performing contractions within each
pair, we get the map
\qq
\lrbc{
\botvV
\xMvr } \otimes
\lrbc{
\boteE
\xtMe}\xrightarrow{\ftot}\ICbphitot,
\label{2.9}
\qqq
where
\qq
\bphiS = \bphiv{1},\ldots,\bphiv{\xNS},
\label{3.8}
\qqq
%
%
and $\xNS$ is the number of connected components $\wsSi$ of
$\wsS$.
If we consider the Lazaroiu connection holonomy along an edge $\xe$ to be the
element of $\xtMe$, then the Wilson network contribution $\WilGa$
can be expressed as the contraction map\rx{2.9} applied to the
tensor product
of all the Lazaroiu holonomies and all the vertex operators
$\xOvr$:
\qq
\WilGa =
\ftot\lrbc{
\lrBigc{
\botvV
\xOvr}\otimes\lrBigc{
\boteE
\Pexp\int_\xe\xAe }
}.
\label{2.9x}
\qqq
Clearly, this expression is a generalization of the \st\ \rx{1.8},
if we assume that the \smgr\
$\xGa$ in the latter case consists of a disjoint union of cycles
formed by the \smeds\ $[\xt_{\xxj-1},\xt_\xxj]$.

\subsection{Operators at the internal points of the \smeds}
The \ntw\ structure of the boundary Wilson lines of the \wsf\
forces us to always include the \smvr\ operators in any correlator
on $\wsfSG$. Moreover, note that the operator spaces $\xHvr$
generally do not contain canonical elements, so there is no
special choice for $\xOvr$.

In addition to the required operators at the \smvrs, the correlators may also
include the optional operators at the internal points of the
\smeds\ $\xe$ and at the internal points of the \ws\ components
$\wsSi$.

Let $\xP$ be an internal point of a \smed\ $\xe$. In order to
describe its space of \locops, we can simply declare it to be a new
\smvr, thus breaking the edge $\xe$ into two consecutive edges.
Then the space of the operators can be presented either in the
form\rx{1.8x2} or in the form\rx{1.8x3}, in which we substitute
$\xMlz=\xMlo=\xMe$. Note that the space $\Ext(\xMe,\xMe)$ has a
canonical element, which is the identity map.

\subsection{\Jca\ and the \locops\ of the \bulk}
The description of the operator spaces for the internal points of
the components $\wsSi$ comes from the \xTLGTs\ on closed surfaces.
Namely, for a \xTLGT\ $\thLGpW$
\qq
\xHP = \ICphi/\iddW,
\label{2.12}
\qqq
where $\bdel\xW$ is the ideal of $\ICphi$ generated by the first
partial derivatives $\deli\xW$ (indeed, $\xHP$ must be the
cohomology of the \BRST\ operator acting according to\rx{1.2} on
the algebra of the fields). The $\ICphi$-module $\xHP$ has an
algebra structure and is called the \Jca\ of $\xW$, so we will also denote
it as $\JW$.

The operator product expansion arguments show that for any \wsf\
component $\wsSi$, whose boundary passes through a \smvr\ $\xv$,
the operator space $\xHvr$ must be a module over the \Jcas\
$\JWi$. The space $\xHvr$ by its definition\rx{2.7} is already a
module over $\ICphii$, so we have to check that if
$\xOvr\in\ker\xDvr$, then
\qq
(\del\xWi/\del\philvv{i}{j})\,\xOvr\in\im\xDvr.
\label{2.13}
\qqq
The proof is based on a general relation
\qq
\{\delj\xD,\xD\} = \delj\xW,
\label{2.14}
\qqq
which follows from \ex{1.4x} by taking the derivative $\delj$ of
both sides. Now
let $\xez$ be an \adjacent\ edge of $\xv$, which is a
part of the boundary $\del\wsSi$.
Suppose that all edges \adjacent\ to $\xv$ go out of $\xv$. Then $\xDvr = \sum_{\xe\in\xIvr}
\xDe$. Since $\{\xDe,\xDez\} = 0$ for all $\xe\neq\xez$,
\ex{2.14} implies that
\qq
\{\delj\xDe,\xDvr\} = \{\delj\xDe,\xDe\} = \delj\xWe = \delj\xWi,
\label{2.15}
\qqq
the latter equality following from the last equation of\rx{2.1}.
Thus
\qq
\delj\xWe\,\xOvr = \{\delj\xDe,\xDvr\}\,\xOvr =
(\delj\xDe)\,\xDvr\,\xOvr + \xDvr\,(\delj\xDe)\,\xOvr.
\label{2.16}
\qqq
Since $\xOvr\in\ker\xDvr$, the first term in the \rhs of this
formula is zero. The second term belongs to $\im\xDvr$, and this
proves our assertion.

\subsection{Topological \LG\ path integral}

Finally, we combine all the data into the path integral, which
represents the correlator of the operators $\xOvr$ at the \sgr\ vertices
and the operators $\xOP$ at the internal points of $\wsS$.
The exponential
part of the integrand is simply $\exp\lrbc{\sum_\xi\mLgrTi}$,
whereas the operators $\xOP$ and the \Wntw\ contribution
$\WilGa$ provide the \preexp\ factors:
\qq
\corrOPOvSG & = \int \exp\lrbc{\sum_\xi\mLgrTi}
\lrbc{\prod_{\xP}\xOP}\WilGa\;
\cD\bphitot\cD\boeta_\tot\cD\btheta_\tot\cD\brho_\tot,
\label{2.10}
\qqq
where $(\mathrm{field})_\tot$ means all fields of the given type
from all the components $\wsSi$.

\subsection{An example of a \xTLGT\ on a \wsf}

Let us now consider a specific example of a \xTLGT\ which can be
put on a \wsf. In other words, we are going to present a set of
\xTLGTs\ $\xTi=\thLGpWi$ and some \mfs\ of the sums of their \sps\ which do not
factor into the tensor products of \mfs\ of the individual \sps\
$\xWi$. This example is inspired by the construction of\cx{Kh1}
and, following\cx{WiVU}, we suggest that it is
the mirror image of the \wsf\ theory presented
in\cx{Ro1}. Also, the constructions of our paper\cx{KR} are based
on a particular case of the \mfs\ described here.

Let us fix a positive integer $\yyN$ and a complex parameter
$\yya$. Following\cx{WiVU}, for $1\leq \yym\leq \yyN-1$ we consider the
polynomial
\qq
\yyciphit = 1 + \sum_{\yyj=1}^\yym \philij\,\yyt^{\yyj},\qquad
\bphim = (\philij\,|\,1\leq \yyj\leq \yym)
\label{2.20}
\qqq
and the expansion of its logarithm in power series of $\yyt$:
\qq
\ln\yyciphit = \syyjoi \yycmjphim\,\yyt^{\yyj}.
\label{2.22}
\qqq
We set
\qq
\xWm(\bphim) = (-1)^{\yyN+1}\,\yycmNo(\bphim) - \yya\,\philmo,
\label{2.23}
\qqq
thus defining the \xTLGT\ $\xTm=\thLGpWm$ with the target space
$\ICm$. Its \Jca\ coincides with the quantum cohomology algebra of
the complex \Grsm\ $\GrmN$.

The \mfs\ that we are going to use belong to a special class
sometimes called the `Koszul factorizations'. Here is the general
construction. Suppose that a \sp\ $\xW(\bphi)$ factors over
$\ICphi$
\qq
\xW = \yyp\,\yyq,\qquad
\yyp,\yyq\in\ICphi.
\label{2.24}
\qqq
Then there exists a $(1|1)$-dimensional \mf\ of $\xW$ with the
\twdif
\begin{eqnarray}
\xD = \lrbc{
\begin{matrix}
0 & \multicolumn{1}{|c}{\yyq}
\\
\cline{1-2}
\yyp &\multicolumn{1}{|c}{0}
\end{matrix}
}.
\label{2.25}
\end{eqnarray}
We denote this \mf\ by $\Kmf{\yyp}{\yyq}$. If $\xW(\bphi)$ can be presented as a sum of products
\qq
\xW = \sum_{\yyj=1}^\yyn\yyp_\yyj\,\yyq_\yyj,\qquad\byyp,\byyq\in\ICphi,
\label{2.26}
\qqq
then there is a $(2^{\yyn-1}|2^{\yyn-1})$-dimensional
factorization of $\xW$
\qq
\Kmf{\byyp}{\byyq} = \bigotimes_{\yyj=1}^{\yyn}
\Kmf{\yypj}{\yyqj}.
\label{2.27}
\qqq

Now we come back to the potentials\rx{2.23}. For a list of integer
numbers
$\yybm$ such that
\qq
\syyi\yymi = \yyN,
\label{2.27x1}
\qqq
we are
going to construct a Koszul \mf\ of the \sp\
\qq
\xWbm=\syyi\xWmi
\label{2.27x2}
\qqq
by presenting it in the
form\rx{2.26}. Consider the polynomial $\xWN(\tbyyp)$ as defined by
\eex{2.20}--\rx{2.23}, in which the variables $\bphi$ are replaced
by the variables $\tbyyp=\xlist{\yypj}{1\leq\yyj\leq\yyN}$.
The equation
\qq
\prod_\yyi \yycmi(\bphimi;\yyt) =1 + \sum_{\yyj=1}^\yyN \yypj(\bphi)\,\yyt^\yyj
\label{2.28}
\qqq
defines $\tbyyp$ as polynomial functions of all variables
$\bphi = \xlist{\bphimi}{\yyi}$. In particular,
\qq
\yypo = \sum_{\yyi}\philmio,
\label{2.30}
\qqq
and it is easy to verify that
\qq
\xWbm(\bphi) = \xWN(\tbyyp(\bphi)).
\label{2.31}
\qqq

Consider again the polynomial $\xWN(\tbyyp)$. If we assign degrees
to the variables $\tbyyp$ as $\deg\yypj = \yyj$, then $\xWN$ is a
homogeneous polynomial of degree $\yyN+1$. Therefore, each monomial
of $\xWN(\byyp)$ is proportional to at least one variable
$\byyp=\xlist{\yypj}{1\leq\yyj\leq (\yyN+1)/2}$ and we can present
$\xWN$ as a sum of products
\qq
\xWN(\tbyyp) = \sjoNot
\yypj\;\yyrj(\tbyyp).
\label{2.32}
\qqq
If we recall that \ex{2.28} turns $\tbyyp$ into polynomials of $\bphi$ and
define the new polynomials $\byyq(\bphi)$ as
\qq
\yyqj(\bphi) = \yyrj(\tbyyp(\bphi)),
\label{2.33}
\qqq
then, according to \ex{2.31},
\qq
\xWbm(\bphi) = \sjoNot \yypj(\bphi)\,\yyqj(\bphi),
\label{2.34}
\qqq
and the \sp\ $\xWbm$ has a \mf\ $\Kmf{\byyp}{\byyq}$. Although the
presentation of $\xWN$ as a sum of products\rx{2.32} is not unique, all these
presentations lead to isomorphic Koszul \mfs. We expect them
to be the mirror images of the special \lgsms\ introduced
in\cx{Kh1} and\cx{Ro1}: a \lgsm\ of the cross-product of the
complex \Grmns\ $\GrmiN$ is defined by the condition that the
subspaces $\IC^{\yymi}\subset\IC^{\yyN}$ provide an orthogonal decomposition of
$\IC^{\yyN}$.

It is interesting to note a resemblance between these \xTLGTs\ and
the representation theory of $\SUN$. If $\yyV$ denotes the
fundamental $\yyN$-dimensional representation of $\SUN$, then the
critical points of a \sp\ $\xWm$ correspond to the weights
of the fundamental representation $\bigwedge^{\yym}\yyV$, a \mf\
$\Kmf{\byyp}{\byyq}$ corresponds to the invariant element in the
tensor product $\bigotimes_{\yyi}\bigwedge^{\yymi}\yyV$ and for a
\lcgr\ $\xgm$ the dimension of the space $\xHgm$ equals the result
of the contraction of the Clebsch-Gordan tensors placed at its
vertices. This correspondence is at the heart of the
categorification construction of\cx{KR}.

\section{Formulas for the correlators}

\subsection{Correlators on a closed surface}
The formula for the correlator of a \xTLGT\ on a closed surface
was derived by Vafa in\cx{V}. Let us define a `\Frt'
map $\ICphi\xrightarrow{\TrW}\IC$ by the formula
\qq
\TrW(\xO) = \frac{1}{(\tpii)^{\xm}}
\oint
\frac{\xO(\bphi)\;d\phi_1\cdots
d\phi_{\xm}}{\delv{1}\xW\cdots\delv{\xm}\xW},
\label{3.1}
\qqq
where the variables $\bphi$ are integrated over the contours
which encircle all critical points of $\xW$.
Let $\xcP$ be a finite set of \emph{\punctures} (marked points) on a
closed surface $\wsS$ of genus $\xg$. According to \cx{V},
the correlator of the operators $\xOP\in\ICphi$ placed at
the punctures $\xP\in\xcP$, is
%
\qq
\corri{\pxcPOP}{\wsS}
 = \TrW\lrbc{\lrbc{\dijW}^\xg
\pxcPOPbphi
}.
\label{3.2}
\qqq
This formula indicates that the \Frt\rx{3.1} computes
the correlator of the operators on a sphere $\St$, while the
factors $\dijW$ represent the `effective contributions' of the
handles: if we imagine that $\wsS$ is a sphere with $\xg$ tori
attached to it by thin tubes, then these tori can be equivalently
replaced by the operators $\dijW$ placed at the points where the
tubes join the sphere.


An important property of the trace\rx{3.1} is that it annihilates
the ideal $\iddW$: for any $\xO\in\ICphi$
\qq
\TrW(\deli\xW\,\xO)=0.
\label{3.3}
\qqq
This is consistent
with the fact that the space of \locops\ is the quotient\rx{2.12}.

\subsection{Correlators on a surface with a boundary}

Suppose that the boundary of the \ws\ $\wsS$ of genus $\xg$ is a union of disjoint
circles\rx{1.6x}, each circle $\xCi$ is split into $\xxnk$ segments
$\xekl$, a \mf\ $\mmfMDWl{\ined}$ is assigned to each segment and
an operator $\xOkl$ is placed at the junction of the segments
$\xev{\ined}$ and $\xev{\xxi,\xxl+1}$. We also place some
operators $\xOP$, $\xP\in\xcP$ at the internal points of $\wsS$. In order to
write a simple expression for the resulting correlator, we have to
introduce a general notation.
For a \mf\
$\mmfMDW$, Kapustin and Li\cx{KL2} define an operator $\dlDw\in\End(\xM)$ by the formula
\qq
\dlDw = \frac{1}{\xm !}\sum_{\sigma\in \rSm } (-1)^{\sign{\sigma}}
\delsv{1}\xD\cdots\delsv{\xm}\xD,
\label{3.4}
\qqq
where $\rSm$ is the symmetric group of $\xm$ elements, $\xm$ being
the dimension of the target space of the \xTLGT\ $\thLGpW$.
Now to a circle $\xCi$, which is a part of the boundary $\dwsS$,
we associate an element $\xOCi\in\ICphi$ defined by the formula
\qq
\xOCi = \STr_{\xMlo}\dlDkow\,\xOv{\xxi,1}\cdots\xOv{\xxi,\xxnk}.
\label{3.5}
\qqq
Note the similarity between the expressions\rx{3.5} and\rx{1.8}:
the former is obtained from the latter by replacing all holonomies
with the identity operators except the first one, which is
replaced by $\dlDkow$.

A.~Kapustin and Y.~Li\cx{KL2} derived the formula for the
correlator of the boundary and \bulk\ operators:
%
\qq
\corri{
\pxcPOP
\prod_{\xxi,\xxl}\xOkl}{\wsS} =
\TrW\lrbc{\lrbc{\dijW}^\xg
\pxcPOPbphi
\prod_\xxi\xOCi}.
\label{3.6}
\qqq
By comparing this formula with \ex{3.2}, we see that the factors
$\xOCi$ represent
the \bstateops\ corresponding to
the boundary
components $\xCi$: the correlator does not change if $\xCi$ is contracted to a point
and the operator $\xOCi$ is placed at that point.

Following\cx{KL2}, let us verify directly (without using the path integral arguments)
that the correlator formula\rx{3.6} satisfies two properties related to the \BRST\
invariance. First of all, if one of the boundary operators $\xOkl$ is \BRST-exact
($\xOkl = \scommm{\xD}{\xOkl\p}$),
then the correlator\rx{3.6} is
zero,
since we can move the operator $\xD$ around the super-trace
expression\rx{3.5}: all the operators $\xO$
commute with $\xD$, while $\{\delj\xD,\xD\}=\delj\xW$ and a
term proportional to $\delj\xW$ is annihilated by the
\Frt.

Second, we could insert the operator $\dlDw$ at any place in
the product of the operators $\xOv{\xxi,1}\cdots\xOv{\xxi,\xxnk}$
in \ex{3.5}: the \rhs of \ex{3.6} would not change.
Indeed,
if for some value of $\xxl$
\qq
\xOkl = \scommm{\delj\xD}{\xOkl\p},\text{ while }
\scommm{\xD}{\xOkl\p} = 0,
\label{3.7}
\qqq
then the \rhs of \ex{3.6} is zero (take the derivative $\delj$ of
the second equation of\rx{3.7} and use the already established fact that
$\xD$-commutators annihilate the \rhs of \ex{3.6}).

\subsection{Correlators on a \wsf}
Now we consider a correlator on a \wsf\ $\wsfSG$.
First, we present a formula and then comment on its
derivation.

%
%
%
We define an operator $\xOGa\in\ICbphiS$ which is the analog of $\xOCi$
from \ex{3.5} and which represents
the \bstateop\
contribution of the
\Wntw. For each connected component $\Cij$ of the boundary
$\del\wsSi$ we choose a \smed\ among the edges to which $\Cij$ is
glued, and we denote that edge as $\eij$. Then we introduce the
operators\rx{3.4}
\qq
\dlDij\in\End(\xMij) 
,\qquad
\dlDij =
\sum_{\sigma\in \rSmi } (-1)^{\sign{\sigma}}
\del_{\philvv{\xi}{\sigma(1)}}\xDv{\eij}\cdots\del_{\philvv{\xi}{\sigma(\xm)}}\xDv{\eij},
\label{3.11}
\qqq
where $\philvv{\xi}{1},\ldots,\philvv{\xi}{\xmi}$ are the bosonic
fields of the \xTLGT\ $\thLGpWi$ assigned to the connected
component $\wsSi$ of $\wsS$.

To every \smed\ $\xe$ we assign an operator
\qq
\xOe = \prod_{(\xij):\,\xe=\eij}\dlDij
\label{3.12}
\qqq
(if $\xe=\eij$ for more than one combination $(\xij)$, then we choose any
order of the operators in the product\rx{3.12}; if $\xe$ never
appears as $\xij$, then the corresponding operator $\xOe$ is the
identity). Similarly to \ex{2.9x}, we define
\qq
\xOGa =
\ftot\lrbc{
\lrBigc{
\botvV
\xOvr}\otimes\lrBigc{
\boteE
\xOe }
}
\label{3.13}
\qqq
and the correlator\rx{2.10} is expressed as
\qq
\corrOPOvSG = \TrS\lrbc{\xOGa
\pxcPOP
\prodiNSdet},
\label{3.14}
\qqq
where
\qq
\xmap{\TrS}{\ICbphiS}{\IC},\qquad \TrS =
\Trv{\xWv{1}}\cdots\Trv{\xWv{\xNS}}
\label{3.9}
\qqq
(\cf \ex{3.6}).


Now let us comment on the derivation of this correlator formula. Note that the original Vafa's
formula\rx{3.2} was derived in the assumption that the critical points of the \sp\
$\xW$ are non-degenerate. Then the \BRST-invariance of the theory
guarantees that the correlator is a sum of the contributions of the
individual critical points and at each point the \sp\ $\xW$ can be
replaced by its quadratic part.

Kapustin and Li derived their formula under
the same assumption, although it is harder to justify in their
case: $\xW$ can be perturbed in order to make its critical points
non-degenerate, but it is not clear whether \mfs\ can be
deformed together with it.

Kapustin and Li derived the correlator for the
disk \ws. This is sufficient in order to establish the state
operator
corresponding to the boundary, and their general formula would
follow from Vafa's formula\rx{3.2}. We use the same
approach. Namely, it would be sufficient to derive \ex{3.14} under the
assumption that the surface $\wsS$ is a union of
disjoint disks. Then the computation of the path integral\rx{2.10}
that leads to \ex{3.14} is exactly the same as in\cx{KL2}.
The only minor novelty is that it may happen that a \smed\ $\xe$
is assigned to two (or more) different disks $\wsSi$ and $\wsSj$
($i\neq j$). Then it bears their operators $\dlDi$ and $\dlDj$ (we
left only one index in their notation, since $\wsSi$ and $\wsSj$
have only one boundary component).
The path integration over the fermionic fields leaves the
derivatives $\dlfikDe$ ($1\leq k\leq \xmi$) and $\dlfjlDe$ ($1\leq
l\leq \xmj$), which enter in the expressions\rx{3.11}, intermixed.
However, they can still be pulled apart into the operators
$\dlDi$ and $\dlDj$, because $\dlfikDe$ and $\dlfjlDe$ anti-commute
up to a \BRST-closed operator. Indeed, since the \sp\ $\xWe$ is a
sum\rx{2.1} of the individual \sps, each depending on its own set
of fields, then $\dlfik\dlfjl\xWe=0$. Hence, if we apply
$\dlfik\dlfjl$ to both sides of the relation $\xDe^2=\xWe$, then
we find that
\qq
\{\dlfikDe,\dlfjlDe\} = - \{\xDe,\dlfik\dlfjl\xDe\}.
\label{3.14x}
\qqq

\section{Gluing formulas}
\subsection{Gluing of a \ynod\ \ws}
The gluing property of the correlators is an important feature of
general \QFTs\ and of topological theories, in particular. Let us
quickly review the gluing rules of a \ynod\ topological \QFT.

For a point $\xP\in\wsS$, let $\yNP$ denote the intersection between
$\wsS$ and a small sphere centered at $\xP$. We will call $\yNP$
the \emph{\lcspsc} of $\xP$. In the context of a string theory
$\yNP$ is simply called a string.
If $\xP$ is an internal point of $\wsS$, then $\yNP$ is a circle (closed string),
and if $\xP$ is a point at the boundary $\dwsS$, then $\yNP$ is a
segment (half-circle, or open string). The endpoints of the segment come from the intersection
of the small sphere and the boundary $\dwsS$, so they are `decorated' with the \TQFT\ boundary conditions
at $\dwsS$. If $\xP$ is a point at the junction of two different
boundary conditions, then the decorations at its endpoints are
also different. The segment is oriented, its orientation being induced by the orientation of
$\wsS$. For a \dlcspsc\ $\xgm$ we define its \dual\ $\xgms$ to be
the same as $\xgm$ but with the opposite orientation. Obviously, a
circle is self-\dual.

The space of the \TQFT\
states corresponding to $\yNP$ coincides with the space
$\xHP$ of the local operators that can be inserted at $\xP$.

Suppose that for two points $\xPo,\xPt\in\wsS$, their \dlcspscs\ are
\dual\ to each other: $\xgmos=\xgmt$. Then the spaces $\xHo$ and
$\xHt$ are also \dual. In order to define the duality pairing
between them, we consider the \ws\ $\xBot =\xUo\#\xUt$
constructed by gluing together small neighborhoods $\xUo$ and
$\xUt$ of $\xPo$ and $\xPt$ over the boundaries $\xgmo\sim\xgmt$
identified with opposite orientations. If $\xgmo$ is a circle,
then $\xBot$ is a 2-sphere, and if $\xgmo$ is a segment, then
$\xBot$ is a disk. The pairing is defined by the correlator on
$\xBot$:
\qq
\dprPpv{\xOo}{\xOt} = \corri{\xOo\,\xOt}{\xBot},\qquad
\xOo\in\xHo,\;\xOt\in\xHt.
\label{4.2}
\qqq
As a result, there is a canonical dual element
\qq
\yIot\in\xH^{\ast}_{1}\otimes\xH^{\ast}_{2},
\label{4.3}
\qqq
defined by the relation
\qq
(\yIot,\xOo\otimes\xOt) = \dprPpv{\xOo}{\xOt}.
\label{4.4}
\qqq
We will need the inverse element
\qq
\yIiot\in\xHo\otimes\xHt.
\label{4.4x}
\qqq

Let us cut small neighborhoods $\xUo$ and $\xUt$ from the
\ws\
$\wsS$ and glue (that is, identify) the boundaries $\xgmo$ and $\xgmt$ of the cuts in
such a way that their orientations are opposite. Denote the
resulting oriented manifold as $\wsSp$,
%
%
then according to the gluing property of a \TQFT,
%
\qq
\corri{\pxcPOP}{\wsSp} = \corri{\yIiot\pxcPOP}{\wsS}.
\label{4.5}
\qqq
A more `pedestrian' way to formulate the same property is to
introduce a basis of operators $\xOj\in\xHo$ and a dual basis
$\xOsj\in\xHt$ so that $\dprPpv{\xOj}{\xOsv{\xxj\p}} =
\delta_{\xxj,\xxj\p}$. Then
\qq
\corri{\pxcPOP}{\wsSp} = \sum_{\xxj}\corri{\xOj\,\xOsj\pxcPOP}{\wsS}.
\label{4.5x}
\qqq

\subsection{Complete space gluing}
\label{s5.2}
Let us check how the general gluing formula\rx{4.5} works for a
\xTLGT\ on a \wsf.

Let $\xP$ be a point on a \wsf\ $\wsfSG$. We call its \lcspsc\ a
\emph{\lcgr} and denote it as $\xgmP$. If $\xP$ is an internal
point of $\wsS$, then $\xgmP$ is a circle. If $\xP$ is an internal
point of a \smed\ $\xe$, then $\xgmP$ is a graph with two vertices
connected by multiple edges, each edge corresponding to a strip of
a component $\wsSi$ attached to $\xe$. If $\xP$ is a \smvr\ $\xv$, then
$\xgmP$ is a graph, whose vertices correspond to the \smeds\
\adjacent\ to $\xv$ and whose edges correspond to the strips of the
components $\wsSi$, which pass through $\xv$. In fact, the
vertex-edge and edge-surface
correspondence between $\xgmP$ and $\wsfSG$ works for all three
types of points $\xP$.
The orientation of the components $\wsSi$ induces the orientation
of the corresponding edges of $\xgmP$.
A \snghb\ $\xUP$ of $\xP$ in $\wsfSG$ can be restored from
its \lcgr\ $\xgmP$, because $\xUP$ is the cone of $\xgmP$:
\qq
\xUP 
= \CngmP.
\label{4.6}
\qqq

Generally speaking, a \lcgr\ $\xgm$ is just a graph. A \emph{\dcr} \lcgr\ means
the following.
To an oriented edge $\yed$ of $\xgm$ we assign
a \xTLGT\ $\thLGyeW$ in such a way that if $\yed$ and $\yeds$
represent the same edge with opposite orientations, then they are
assigned the conjugated theories. To a vertex $\yvr$ of $\xgm$ we
associate a matrix factorization $\mmfMDWyed$, such that
\qq
\xWyvr = \sum_{\yed\in\silgyvr}\xWye,
\label{4.6y}
\qqq
where $\silgyvr$ is the set of edges of $\xgm$, which are attached
to $\yvr$ (we assume that they are oriented away from $\yvr$).

For a \dlcgr\ $\xgm$ consider the \mf\ $\mmfMDWgm$, which is the tensor product of all the \mfs\ of
its vertices
\qq
\mmfMDWgm = \bigotimes_{\yvr}\mmfMDWyed.
\label{4.6y1x}
\qqq
Obviously
\qq
\xWgm=0,
\label{4.6y1}
\qqq
so $\xDgm^2 = 0$ and we denote
its cohomology as
\qq
\xHgm = \ker\xDgm/\im\xDgm.
\label{4.6y2}
\qqq

If a \wsf\ $\wsfSG$ is \dcr, then for any $\xP\in\wsfSG$ its
\lcgr\ $\xgmP$ is also \dcr: a \xTLGT\ of an edge $\yed$ is the
theory of the corresponding component $\wsSi$ and a \mf\ of a
vertex $\yvr$ is the \mf\ of the corresponding \smed, if it is
oriented away from $\xP$, or the conjugated \mf\ otherwise. Then it
is easy to see that
\qq
\xHgmP = \xHP,
\label{4.6y3}
\qqq
that is, the space of operators at a point $\xP$ is determined by
its \dlcgr\ $\xgmP$.

For a \dlcgr\ $\xgm$ we define its \dual\ graph $\xgms$ to be the same graph as $\xgm$,
except that it is \dcr\ with the conjugate \xTLGTs\ and with the dual
\mfs.

Consider a suspension $\Suspgm$ of a \lcgr\ $\xgm$: by definition
it is constructed by gluing together the cones $\Cngm$ and
$\Cnv{\xgms}$ along their common boundary $\xgm$. $\Suspgm$ has a
structure of a \wsf: its \smgr\ consists of two vertices $\xvro$ and $\xvrt$, which are
the vertices of the cones, the suspensions of the vertices of
$\xgm$ form the edges that connect $\xvro$ and $\xvrt$, and the
2-dimensional components $\wsSi$ are the suspensions of the edges
of $\xgm$. Obviously, $\xgmv{\xvro}=\xgm$ and
$\xgmv{\xvrt}=\xgms$.

If a \lcgr\ $\xgm$ is \dcr, then a \xTLGT\ is defined on its
suspension $\Suspgm$. The correlators of this theory provide a
pairing between the spaces $\xHgm$ and $\xHgms$
\qq
\dprv{\xO}{\xOp}{} = \corri{\xO\,\xOp}{\Suspgm}.
\label{4.7}
\qqq
This pairing determines an inverse canonical element
$\xIgs\in\xHgm\otimes\xHgms$.

Suppose that for two points
$\xPo,\xPt\in\wsfSG$ their \lcgrs\ are \dual: $\xgmos = \xgmt$.
Then we can cut their small neighborhoods from the \wsf\ $\wsfSG$ and
glue the cut borders together, thus forming a new \wsf\ $\wsfSGp$.
If a \xTLGT\ is defined on $\wsfSGp$, then it induces a \xTLGT\ on
$\wsfSG$ and the correlators of both theories are related by the
gluing formula
\qq
\corrOPOvSGp = \corrIOPOvSG,
\label{4.8}
\qqq
where $\xcPp$ and $\xcVp$ are the punctures and \smvrs\ of
$\wsfSGp$.
Thus the gluing property of a \xTLGT\ on a \wsf\ is very similar
to the gluing property\rx{4.5} on a usual \ws.

\subsection{A \xTLGT\ on a \wsf\ as a \tcategory}

The graph structure of \wsf\ \lcspscs\ permits more complicated
types of gluing than those described in the previous subsection,
when two \dual\ \lcspscs\ are glued together. These new types of
gluing can be arranged into the mathematical structure known as a
\tcategory.

The usual \ocategory\ structure of a \TQFT\ on a 2-dimensional \ws\
comes from the composition property of \tramps. Consider a \ws\
$\wsS$ with a finite \puncture\ set $\xcP$ and two special
\punctures\
$\xPo$ and $\xPt$ with \lcgrs\ $\xgmo$, $\xgmt$. If we choose the operators
at the punctures of $\xcP$,
then the correlator on $\wsS$
defines a \tramp\
\qq
\xHgmo\xrightarrow{\xAgost}\xHgmst
\label{5.1}
\qqq
by the formula
\qq
\yIgmts(\xAgost(\xOo),\xOt) = \corri{\xOo\xOt\pxcPOP}{\wsS}
\quad\text{for any}\quad\xOo\in\xHgmo,\;\xOt\in\xHgmt.
\label{5.2}
\qqq
%
Let $\wsSot$
denote the result of cutting \snghbs\ of $\xPo$ and $\xPt$ from
$\wsSot$. If we have another surface $\wsSp$ with special
\punctures\ $\xPth$ and $\xPf$ such that $\xgmst=\xgmth$,
then we can glue the boundary components $\xgmt$ and $\xgmth$ of
$\wsSot$ and $\wsSpthf$ together to form a new \ws\ with boundary
$\wsSppof$. The \tramp\ of $\wsSpp$ is given by the composition of
\tramps
%
%
\qq
\xAgosf = \xAgthsf\,\xAgost,
\label{5.3}
\qqq
and this formula corresponds to the gluing formula\rx{4.5}
formulated for $\xPt$ and $\xPth$.

The composition
property\rx{5.3} extends verbatim to the \wsfs. In the
\foam\
case, however,
there is an important generalization. Namely, the surface or the
\wsf\ $\wsSot$ presented a cobordism between two closed \spsc\ (be it 1-manifolds or graphs).
Now we are going to
consider cobordisms between the spaces that have boundaries.

Let us take a \lcgr\ $\xgm$ and make cuts across some of its
edges,
so that $\xgm$ splits into two disconnected \emph{\plcgrs} $\aplgo$ and
$\aplgt$: $\xgm=\aplgo\#\aplgt$. The \plcgrs\ have special univalent vertices at the cuts:
we call them \emph{\bvertices}, and their \adjacent\ edges are called \emph{\legs}. We think
of the
\bvertices\ as the \boundary\ of a \plcgr.

The \plcgrs\ inherit the
decorations of $\xgm$, except that their \bvertices\ are not assigned
\mfs.
To a \dplgr\ $\aplg$ we associate a \mf\
$\mmfMDWa$ which is the tensor product of all the \mfs\ of
its non-\boundary\ vertices. However this time instead of \ex{4.6y1} we have
\qq
\xWa = \sum_{\edLa} \xWye,
\label{5.4}
\qqq
where $\La$ is the set of legs of $\aplg$, and we assume that the
legs are oriented away from the \bvertices.

Let $\bphia=\xlist{\bphiye}{\edLa}$ be the list of all variables
of the \legs\ of a \plcgr\ $\aplg$, and let $\xRa=\ICw{\bphia}$ be
their polynomial ring. Obviously,
\qq
\xRa = \bigotimes_{\edLa}\ICbphiye.
\label{5.5}
\qqq
Since the \sp\ $\xWa$ of \ex{5.4} depends only on the `external'
variables $\bphia$, from now on we will consider $\mmfMDWa$ to be a \mf\ over
the ring $\xRa$. However this poses a problem: if we ignore the
`internal' variables of $\aplg$, then the module
$\xMa$ is infinite-dimensional as a module over $\xRa$. Indeed,
the multiplication by the powers of an internal variable now produces
an infinite sequence of linearly-independent elements of $\xMa$.
In order to resolve this problem, we can contract $\xMa$
homotopically to a finite-dimensional $\xRa$-module.
Here is the
relevant definition: two \mfs\ $\mmfMDWxi$, $i=1,2$ are considered \emph{homotopically
equivalent} over the polynomial ring $\xR\ni\xW$, if there exist two
$\xR$-linear maps $\yfot$, $\yfto$
\qq
\xMlo\xrightarrow{\yfot}\xMlt\xrightarrow{\yfto}\xMlo
\label{5.6}
\qqq
commuting with the \twdif\ $\xD$, such that the compositions
$\yfto\yfot\in\EndR(\xMlo)$ and $\yfot\yfto\in\EndR(\xMlt)$ are
\BRST-equivalent to the identity maps. We showed in\cx{KR} that
under some mild assumptions an infinite rank
\mf\ is homotopically equivalent to a finite rank one.

If a \dlcgr\ is split: $\xgm = \aplgo\#\aplgt$, then
according to \ex{4.6y1x}
\qq
\xMgm = \xMao\otR\xMat,\qquad\xR = \xRv{\aplgo} = \xRv{\aplgt},
\label{5.7}
\qqq
and we used the notation $\otR$ in order to emphasize that the
tensor product is taken over the polynomial ring of all \leg\
variables of $\aplgo$ (or, equivalently, $\aplgt$). If we replace
the $\xR$-modules $\xMao$ and $\xMat$ in \ex{5.7} by their
finite-dimensional homotopic equivalents, then the module $\xMgm$
will change, but its $\xD$ cohomology $\xHgm$ will stay the same.
Therefore we will use the same notation $\xMa$ for the whole
homotopy equivalence class of $\xRa$-modules related to a \dplgr\
$\aplg$.

Let $\wsfSG$ be a \dwsf\ with a puncture set $\xcP$. Let us pick a
vertex $\xv\in\xcV$ with a \lcgr\ $\xgm$ and denote $\xcVp = \xcV\setminus\{\xv\}$.
A choice of the operators at the punctures of $\xcP$ and at the vertices
of $\xcVp$ determines an element $\xAgm\in\xHgms$ by the formula
\qq
\yIgsg(\xO,\xAgm) =\corrSG{\xO\pxcPOP\pxcVpOvp}\quad\text{for
any}\quad\xO\in\xHgm.
\label{5.8}
\qqq
Suppose that $\xgm$ splits: $\xgm = \aplgo\#\aplgt$. Then, in view
of \ex{5.7},
\qq
\xMgms = \xMaso\otR\xMast = \HomR(\xMao,\xMast),
\label{5.9}
\qqq
and hence
\qq
\xHgms = \Ext(\xMao,\xMast)
\label{5.10}
\qqq
(\cf the definition\rx{1.8x2}). The latter isomorphism allows us
to translate the element $\xAgm$ of \ex{5.8} into a \tramp\
$\xAaot\in\Ext(\xMao,\xMast)$. This \tramp\ is the analog of the
amplitude\rx{5.1}: $\xAgost$ describes the transition between two
closed \spscs, while $\xAaot$ describes the transition between two
\spscs\ with boundary.

The distinction between the open and closed space transitions has a topological and
an algebraic manifestation.
Topologically, if we cut \snghbs\ of the \punctures\ $\xPo$
and $\xPt$, then the remainder $\xncwsfSGPot$ has two disconnected boundary
components: the `in' space $\xgmo$ and the `out' space $\xgmt$.
The \foam\ topology is more complicated. If we cut out a \snghb\
of the \smvr\ $\xv$, then the boundary of the remainder $\xncwsfSGPvr$ is obviously the \lcgr\
$\xgm$. If we cut $\xgm$ just into $\aplgo$ and $\aplgt$, then these
\plcgrs\ would have common points. However, since we think of
$\aplgo$ and $\aplgt$ as \spscs\ corresponding to different values
of `time', we would like them to be completely separated.
Therefore, rather than slicing the edges
that connect $\aplgo$ and $\aplgt$, we cut out
finite length segments from them. These segments form the
\emph{\tlsc}. Thus the boundary of $\xncwsfSGPvr$ consists
of three, rather than two, pieces: two `space-like' ones (the `in' space $\aplgo$ and the
`out' space $\aplgst$) as well as the \tlsc. The
algebraic consequence of the presence of a \tlsc\ in the boundary of
$\xncwsfSGPvr$ is that the `in' and `out' spaces of states are not just linear spaces over $\IC$, but
rather $\xR$-modules, and the \tramp\ $\xAaot$ is $\xR$-linear.

Since the \wsf\ of the open space transition $\xncwsfSGPvr$ has two types of boundary,
the corresponding \tramp\ $\xAaot$
satisfies two gluing relations. First of all, there is the gluing
associated to the composition of transitions, which is similar to
\ex{5.3}. Consider two \wsfs\ $\wsfSGj$ ($\xj=1,2$) with marked
\smvrs\ $\xvrj$. Suppose that their \lcgrs\ $\xgmj$ can be split
$\xgmj = \aplgj\#\aplgpj$ in such a way that $\aplgst = \aplgpo$.
Then we can glue $\xncwsfSGvro$ and $\xncwsfSGvrt$ along these matching
\plcgrs. Its resulting \tramp\ $\xAaot$ should be the composition
of the elementary ones:
\qq
\xAaot = \xAat\,\xAao.
\label{5.11}
\qqq

One can also glue the \wsfs\ along the \tlscs. Let us describe the corresponding
cutting of a \wsf $\wsfSG$. Suppose that it has two marked vertices $\xvrj$ ($\xj=1,2$), whose \lcgrs\
$\xgmj$ are split: $\xgmj = \aplgpj\#\aplgppj$, and
both $\xgmo$ and $\xgmt$ have $m$ \sl\ points.
Let $\xpjk$ ($1\leq
\xk\leq m$) denote the points at which the legs of $\aplgpj$ and
$\aplgppj$ are joined. Suppose that for all $\xk$, the points
$\xpok$ and $\xptk$ belong to the same connected component
$\wsSik$ of $\wsS$, and we can choose nonintersecting
curves $\cack$ which lie on $\wsSik$ and join the points $\xpok$
and $\xptk$. Finally, suppose that if we make the cuts along all curves
$\cack$, then $\xncwsfSGPvot$ splits into two disconnected pieces
$\xspvotp$, $\xspvotpp$ in such a way that $\xspvotp$ is bound
by $\aplgpo$, $\aplgpt$ and the curves $\cack$, while $\xspvotpp$ is
bound by $\aplgppo$, $\aplgppt$ and the curves $\cack$. Then
$\xspvotp$ and $\xspvotpp$ produce their own \tramps\
$\xAapot$ and $\xAappot$. Their tensor product
\qq
\xMapo\otR\xMappo\xrightarrow{\xAapot\otR\xAappot}
\xMapst\otR\xMappst
\label{5.12}
\qqq
commutes with the \twdif\ $\xD$ and therefore, in view of\rx{5.7},
it defines a map from $\xHgmo$ to $\xHgmst$. Thus, the gluing of
two \wsfs\ $\xspvotp$ and $\xspvotpp$ along their \tlscs\ produces
the formula
\qq
\xAgost = \xAapot\otR\xAappot,
\label{5.13}
\qqq
relating two open-space \tramps\ to one closed-space \tramp. In this gluing formula the usual
composition\rx{5.3} and\rx{5.11} is replaced by the tensor product
over an appropriate ring.

\subsection*{Acknowledgements}
L.R. would like to thank Anton Kapustin for many useful
discussions and explanations regarding 2-dimensional topological
theories and their boundary conditions.

This work was supported by NSF Grants DMS-0104139 and DMS-0196131.


\thebibliography{99}
%
\bibitem{BHLS}I.~Brunner, M.~Herbst, W.~Lerche, B.~Scheuner,
\emph{Landau-Ginzburg realization of open string TFT}, preprint
hep-th/0305133.
%
\bibitem{La1} C.~Lazaroiu, \emph{On the boundary coupling of topological Landau-Ginzburg
models}, preprint hep-th/0312286.
%
\bimnx{KL2}{A.~Kapustin, Y.~Li}{Topological Correlators in Landau-Ginzburg Models with
Boundaries}{Adv. Theor. Math.
Phys.}{7}{2003}{727-749,} hep-th/0305136.
%
\bibitem{Kh1}M.~Khovanov, \emph{sl(3) link homology}, preprint
math.QA/0304375.
%
\bibitem{KR}M.~Khovanov, L.~Rozansky, \emph{Matrix factorizations and link homology}, preprint
math.QA/0401268.
%
\bibitem{Ro1}L.~Rozansky, \emph{Topological A-models on seamed
Riemann surfaces}, preprint hep-th/0305205.
%
\bimn{V}{C.~Vafa}{Topological Landau-Ginzburg models}{Mod. Phys.
Lett}{A6}{1991}{337}
%
\bibitem{WiVU}E.~Witten, \emph{The Verlinde algebra and the
cohomology of the Grassmannian}, In: `Cambridge 1993, Geometry,
topology and physics', pp. 357-442, preprint hep-th/9312104
%
\end{document}